\documentclass{aa}  

\usepackage{graphicx}

\usepackage{amsmath,mathtools}
\usepackage{amssymb}
\usepackage{graphicx}
\usepackage{enumitem}

\usepackage{color}

\usepackage{natbib}
\usepackage[english]{babel}

\usepackage{tabularx}
\usepackage{multirow}
\usepackage{xspace} 
\usepackage{array,booktabs}
\newcolumntype{L}{>{\raggedright\let\newline\\\arraybackslash\hspace{0pt}}X}

\usepackage{braket}

\usepackage[hidelinks]{hyperref}

\newcommand{\dd}{\mathrm{d}}

\newcommand{\mrm}[1]{\mathrm{#1}}
\newcommand{\pow}[1]{\ifmmode{}^{#1}\else ${}^{#1}$\fi}
\newcommand{\HI}{{\text{H\MakeUppercase{\romannumeral 1}}}\xspace}

\newcommand{\Lya}{\ifmmode{\mathrm{Ly}\alpha}\else Ly$\alpha$\xspace\fi}

\newcommand{\e}{\mathrm{e}}

\newcommand{\cm}{\,\ifmmode{{\rm cm}}\else cm\fi}
\newcommand{\ccm}{\,\mathrm{cm}^{-3}}
\newcommand{\ergps}{\,{\rm erg}\,{\rm s}\ifmmode{}^{-1}\else ${}^{-1}$\fi}
\newcommand{\Mpch}{\,{\rm Mpc}\,\ifmmode h^{-1}\else $h^{-1}$\fi}
\newcommand{\snru}{\,\ifmmode{\mathrm{Myr}^{-1}}\else Myr${}^{-1}$\fi}
\newcommand{\kms}{\,\ifmmode{\mathrm{km}\,\mathrm{s}^{-1}}\else km\,s${}^{-1}$\fi}

\def\lsim{~\rlap{$<$}{\lower 1.0ex\hbox{$\sim$}}}

\def\gsim{~\rlap{$>$}{\lower 1.0ex\hbox{$\sim$}}}
\newcommand{\fc}{\relax\ifmmode{f_{\mathrm{c}}}\else  $f_{\mathrm{c}}$\xspace\fi}
\newcommand{\fccrit}{\relax\ifmmode f_{\mathrm{c,\,crit}}\else  $f_{\mathrm{c,\,crit}}$\xspace\fi}
\newcommand{\cl}{{\mathrm{cl}}}

\begin{document}

\title{Resonant line transfer in a fog:\\
Using Lyman-alpha to probe tiny structures in atomic gas}
\titlerunning{Resonant line transfer in a fog}

\author{Max Gronke\inst{1}, Mark Dijkstra\inst{1}, Michael McCourt\inst{2,3}, S. Peng Oh\inst{2}}

   \institute{Institute of Theoretical Astrophysics, University of Oslo, Postboks 1029, 0315 Oslo, Norway\\
              \email{maxbg@astro.uio.no}
         \and
Department of Physics, University of California, Santa Barbara, CA 93106, USA
\and
Hubble fellow
             }

   \date{Draft from \today}

  \abstract
{
Motivated by observational and theoretical work which both suggest very small
scale ($\lesssim 1\,$pc) structure in the circum-galactic medium of galaxies and
in other environments, we study Lyman-$\alpha$ (Ly$\alpha$) radiative transfer
in an extremely clumpy medium with many ``clouds'' of neutral gas along the line
of sight. While previous studies have typically considered radiative transfer
through sightlines intercepting $\lesssim 10$ clumps, we explore the limit of a
very large number of clumps per sightline (up to $f_{\mathrm{c}} \sim 1000$).
Our main finding is that, for covering factors greater than some critical
threshold, a multiphase medium behaves similar to a homogeneous medium in terms
of the emergent Ly$\alpha$ spectrum. The value of this threshold depends on both
the clump column density and on the movement of the clumps. We estimate this
threshold analytically and compare our findings to radiative transfer
simulations with a range of covering factors, clump column densities, radii, and
motions. Our results suggest that \textit{(i)} the success in fitting observed
Ly$\alpha$ spectra using homogeneous ``shell models'' (and the corresponding
failure of multiphase models) hints towards the presence of very small-scale
structure in neutral gas, in agreement within a number of other observations;
and \textit{(ii)} the recurrent problems of reproducing realistic line profiles
from hydrodynamical simulations may be due to their inability to resolve
small-scale structure, which causes simulations to underestimate the effective
covering factor of neutral gas clouds.
}



   \maketitle



\section{Introduction}
Several observations highlight the presence of tiny, unresolved
structure in atomic gas across a wide range of astrophysical
environments.  For instance, the wide, smooth emission lines in quasar
spectra suggest the atomic gas close to the black hole has both a
suprathermal velocity dispersion but low volume filling factor
\citep[e.\,g.][]{Rees1987,1997MNRAS.288.1015A}.  Moreover, studies of
the diffuse gas in the halos of massive galaxies at redshifts
$z\sim{2-3}$ routinely find that these galaxies are filled with tiny
clouds of neutral gas, again with a high covering factor, but with a
low overall volume filling factor
\citep[e.\,g.][]{Rauch1999,Cantalupo2014,Hennawi2015}.  Similar
evidence for tiny-scale structure in neutral gas may be found in
galactic winds and in high-velocity clouds in the Milky Way (see,
e.\,g. \citealt{McCourt2016} for a summary).

The physical origin of these clumps has been investigated recently by
\citet{McCourt2016}, who find that cooling gas clouds are prone to
rapid fragmentation akin to the Jeans instability. They suggest this
fragmentation rapidly ``shatters'' cold gas into tiny cloudlets of a
characteristic size $l \sim 0.1{\,\rm pc} (n / \cm^{-3})^{-1}$, or
equivalently a column density $N_{\mathrm{cloudlet}} \sim
10^{17}\cm^{-2}$.  \citet{McCourt2016} argue that this length scale is
consistent with a number of observational upper limits, but
unfortunately such small scales are extremely difficult to probe
directly in distant objects.  In this paper, we show that radiative
transfer of the resonant Ly$\alpha$ line can indeed probe sub-parsec
scales, even in distant galaxies.

There are several reasons why the \Lya emission line hydrogen is an ideal probe for tiny-scale structures.
As the most prominent transition line of the most abundant element, \Lya is a sensitive probe of neutral gas enabling us to study even the most distant objects in the Universe.
Recently, instruments such as 
\textit{MUSE} \citep{2010SPIE.7735E..08B} reveal the ubiquity of \Lya emission throughout the observable space. In particular, \Lya is used to study our galactic neighborhood \citep{Hayes2015}, galaxies at the peak of cosmic star formation \citep{Barnes2014}, and the later stages of reionization \citep{Dijkstra2014_review}.

Apart from this practical reason, the resonant nature of the \Lya
transition gives \Lya observations a potentially great constraining
power in studying otherwise unresolvable structure.  This is due to
the strong frequency dependence of the neutral hydrogen scattering cross
section which leads to many orders of magnitude of variation in the photon
mean free path.  For instance, in a medium with one neutral hydrogen
atom per $\ccm$, a \Lya photon travels on average only $\sim 1\,$AU
if it is in the core of the line; however, this distance grows by
nearly five orders of magnitude to $\sim 0.5\,$pc if the frequency is
shifted merely five Doppler widths ($\sim{60}$\,km/s) away from line
center.  The mean free is therefore also sensitive to gas motions on
the scale of $\sim(1-100)$\,km/s, providing powerful constraints on
the kinematic properties of galaxies and their surrounding environments.

In this paper, we revisit Ly$\alpha$ radiative transfer through a
simplified clumpy medium consisting of spherical ``clouds'' of neutral
hydrogen embedded in an ionized surrounding medium.  While this setup
has been considered many times before
\citep[e.\,g.][]{Neufeld1991,Hansen2005,Laursen2012,Duval2013,Gronke2016a},
all of these previous studies have focused on a part of the parameter
space with a relatively low ($\lesssim 10$) number of clumps per
sightline.  In light of the observations and the shattering scenario
discussed above, we now consider the limit with many more clouds per
sightline, exploring the full range from $\sim{1}$ to $\sim{1000}$s.
We show that this has tremendous influence on the propagation of
Ly$\alpha$, and we provide simple scaling relations which enable
simple, order-of-magnitude calculations for Ly$\alpha$ radiative
transfer through clumpy media.

Our paper is structured as follows. In Sec.~\ref{sec:analyt-cons-new}, we discuss the problem analytically and estimate the expected results.
In Sec.~\ref{sec:method} we describe briefly our \Lya radiative transfer calculations, and introduce our model. We present the simulation results in Sec.~\ref{sec:results} with particular focus on the spectral shape and the \Lya escape fraction as well as their connections to a corresponding homogeneous medium. We then discuss the results in Sec.~\ref{sec:discussion} before we conclude in Sec.~\ref{sec:conclusion}.

\section{Analytic results}
\label{sec:analyt-cons-new}

We find several distinct regimes for \Lya radiative transfer in multiphase
media, which we summarize in figure~\ref{fig:sketch_regimes}.  In this
section, we describe the physics relevant to each regime and provide
analytic estimates for the boundaries separating them.  Since it will
prove essential for our analysis, we first review some general results
about \Lya escape from a homogeneous slab
(\S~\ref{sec:radi-transf-homog}) before describing radiative transfer
in clumpy medium (\S~\ref{sec:radi-transf-clumpy}).  We confirm these
analytic results numerically in section~\ref{sec:results} using
Monte Carlo radiative transfer simulations.


\newcommand{\heading}[1]{\multicolumn{1}{c}{\textbf{#1}}}

\begin{table*}
  \centering
  \caption{Summary of the regimes found in a static, clumpy medium}

  \begin{tabular}{llllll}
    \toprule
    \heading{Name}
    & \heading{Requirement\tablefootmark{a}}
    & \heading{$N_\cl$\tablefootmark{b}}
    & \heading{$x_{\mrm{p}}$\tablefootmark{c}}
    & \heading{Description} \\

    \midrule
    Free-streaming
    & $\tau_{0,\mrm{total}} \lesssim 1$                         
    & $\propto \fc$ 
    &  $\sim 0$ 
    & Photons stream through clumps
    \\
    \addlinespace[1em]

    Porous         
    & $(\tau_{0,\mrm{total}}\gtrsim 1) \land (\fc \lesssim 1)$  
    & $\sim 0$ 
    & $\sim 0$ 
    & Escape through holes without interaction
    \\
    \addlinespace[1em]

    Random-walk    
    & $(\tau_{0,\mrm{total}}\gtrsim 17/a_v) \land (1 \lesssim \fc
    \lesssim \fccrit^{\mrm{exc}})$   
    & $\propto \fc^2$ & $\sim 0$ 
    & Scatter off clumps until escape 
    \\
    \addlinespace[1em]

    \multirow{2}{*}{Homogeneous~~\Bigg\{}    
    & $(\tau_{0,\mrm{total}}\gtrsim 17/a_v) \land (\fc \gtrsim
    \fccrit^{\mrm{exc}})$ 
    & $\propto \fc$ 
    & $\sim x_{\mrm{esc}}$\tablefootmark{A}
    & Excursion-like escape
    \\
    \addlinespace[1em]

    & $(17/a_v \gtrsim \tau_{0,\mrm{total}}\gtrsim 1) \land (\fc
    \gtrsim \fccrit^{\mrm{exc}})$ 
    & $\propto f_c$ & $\sim x_*$\tablefootmark{B}
    & Escape in a single flight 
    \\

    \bottomrule
    
  \end{tabular}

  \tablefoot{The regimes are described in detail in \S~\ref{sec:radi-transf-clumpy}.\\
    \tablefoottext{a}{Parameter space of regime. Visualized in Fig.~\ref{fig:sketch_regimes}.\\}
    \tablefoottext{b}{Average number of clumps a photon encounters before escape.\\}
    \tablefoottext{c}{Dominant frequency of escape.}\\
    \tablefoottext{A}{Escape frequency from a homogeneous slab (Eq.~\eqref{eq:xesc_adams})}.\\
    \tablefoottext{B}{Boundary between core and wing of the line. $x_*\sim 3.26$ for $T=10^4\,$K.}
  }
  \label{tab:regimes}
\end{table*}

\subsection{Definitions \& notation conventions}
\label{sec:general-conventions}
The basics of \Lya radiative transfer have been described in the literature \citep[e.g., recently in a review by][]{Dijkstra2014_review} and will not be repeated in detail here. Instead, we summarize the most relevant quantities for our present applications.
\begin{itemize}
\item We express the photon's frequency $\nu$ in terms of its Doppler parameter
\begin{equation}
x = \frac{\nu - \nu_0}{\Delta \nu}\;,
\end{equation}
where $\nu_0 \approx 2.47\times 10^{15}\,{\rm s}^{-1}$ is the frequency at line center, and $\Delta\nu_D = v_{\rm th} \nu_0 / c = \sqrt{2 k_B T/m_H} \nu_0 / c$ is the line width due to thermal motions of the atoms.
\item Temperature dependence is expressed through the Voigt parameter
\begin{equation}
a_v = \frac{\Delta\nu_L}{2 \Delta_D} \approx 4.7\times 10^{-4} \left(\frac{T}{10^4\,{\rm K}}\right)^{-1/2}\;.
\end{equation}
Here, $\Delta\nu_L= 9.939\times 10^7\,\mrm{s}^{-1}$ is the natural
(i.\,e., quantum mechanical) line broadening due to the finite lifetime of the transition.
\item The \Lya cross section of neutral hydrogen is 
\begin{align}
\sigma_{\HI}(x, T) =& 
\sigma_0 H(a_v, x) \nonumber \\
=& \frac{\sigma_0 a_v}{\pi} \int\limits_{-\infty}^{\infty}\dd y \frac{e^{-y^2}}{(y - x)^2 + a_v^2}
\end{align}
where $\sigma_0 \approx 5.895\times 10^{-14} (T / 10^4\,\mrm{K})^{-1/2}\cm^{2}$ denotes its value at line center and $H(a_v, x)$
is the Voigt function which can be approximated as $H(a_v,
x)\sim{e^{-x^2}}$ in the core of the line and as $\sim
a_v/(\sqrt{\pi}x^2)$ in the wing of the line.  The transition occurs
at a frequency $x_* \approx 3.26$ for $T = 10^4\,$K. The normalized Voigt distribution
$\phi(x) = H(a_v,x) / \sqrt{\pi}$ represents the probability of a photon in the frequency interval $[x\pm \dd x / 2]$ to interact with an atom.
\item The optical depth per length $d$ is, hence,
\begin{equation}
\tau(x) = \int\limits_0^d\dd s\, \sigma_{\rm HI}(x) n_{\rm HI}(s)
\end{equation}
where $n_{\rm HI}$ denotes the number density of neutral hydrogen atoms. Note that we did not include the contribution of dust in the above expression as its impact is modelled in post-process (see \S~\ref{sec:dust-within-clumps} for details).
\end{itemize}

\subsection{Radiative transfer in a homogeneous slab}
\label{sec:radi-transf-homog}

Since it is crucial for our analytic work below, we briefly review some classical solutions for \Lya radiative transfer in a semi-infinite (that is, only one dimension is finite) slab with half-height $B$ and optical depth $\tau_0$.
 
\Lya escape can be seen as a random-walk in both real space and frequency space, as every scattering event (that is, absorption and quick re-emission from a neutral hydrogen atom) alters the frequency and direction of the \Lya photon. However, due to the large value of $\sigma_{0}$, the mean free path of a photon close to line center is very small ($\lambda_{\mathrm{mfp}}\sim 5.5 \times 10^{-6} \left(n / \cm^{-3}\right)^{-1}\,$pc for $T=10^4\,$K), and most scatterings are spatially close to each other.
Thus, the vast majority of scatterings do not contribute significantly towards the escape of the \Lya photon (at least in optically thick media\footnote{For lower optical depths (when $\tau(x_*)\lesssim 1$), an escape in single flight is possible even in the Doppler core. Such escape occurs via rare scattering events when a photon near line center encounters a fast moving atom, with large velocities perpendicular to the photon's direction. When this photon is re-emitted, it will be far from line center, and if $\tau(x)\approx\tau_{0}{\rm e}^{-x^{2}} < 1$, it can escape (also see \S~\ref{sec:divis-betw-regim}).}). Instead, \citet{Adams1972} found that \Lya photons escape in several consecutive wing-scatterings where the mean free path is significantly enhanced (for instance, $\lambda_{\mathrm{mfp}} \sim 0.48\,$pc at $x=5$ for the above setup). 
The random-walk in frequency space is therefore crucial to the escape
of Ly$\alpha$.
These series of wing-scatterings are referred to as `excursion', and this is thought of as the common way \Lya photons propagate in an astrophysical context. 

To estimate the average displacement per `excursion,' one has to take into account its random walk in frequency. Specifically, a \Lya photon in the wing of the line at frequency $x$ has a slight tendency to return to the core of the line with mean frequency shift per scattering event of $-1/x$ \citep{Osterbrock1962}. This means it will scatter $\sim x^2$ times before returning to the core with a mean free path of $l = B \sigma_0 / (\sigma_{\HI}(x) \tau_0) = B / (H(x) \tau_0)$  
between each scatter. Since an excursion itself can be seen as a random walk, \citet{Adams1972} obtained $d_{\mrm{exc}}= \sqrt{N_{\mrm{sct, exc}}} l = x B / (H(x) \tau_0)$ as mean distance per excursion. Furthermore, by using the wing-approximation $H(x)\sim a_v/(\sqrt{\pi} x^2)$ described above, and setting $d_{\mrm{exc}}= \sqrt{3} B$ \citep[where the factor $\sqrt{3}$ arises due to geometrical considerations,][]{Adams1975}, he obtained
\begin{equation}
x_{\mrm{esc}} = \left(\tau_0 a_v \sqrt{3/\pi}\right)^{1/3} \approx 6.5  \left(\frac{N_\HI}{10^{19}\cm^{-2}} \frac{10^4\,\mathrm{K}}{T} \right)^{1/3}
\label{eq:xesc_adams}
\end{equation}
as an expression for the most likely escape frequency.

\citet{Adams1972} continues to calculate the number of scatterings it takes for a photon to reach a frequency $|x| \ge x_{\mrm{esc}}$ which allows for escape. In an optically thick medium
photons undergo many scatterings and the frequency distribution $J(x)$ is roughly constant\footnote{Like any diffusive process, frequency diffusion can be represented by a Fokker-Planck equation, for which the steady state solution is $J(x)=\mathrm{const}$. This is independent of the form of the frequency diffusion coefficient (and thus independent of the redistribution function).}. Thus, the probability to find an arbitrary photon with frequency in the interval $[x\pm \dd x / 2]$ is $\sim \phi(x) \dd x$ (complete redistribution approximation\footnote{This approximation holds only for $|x|\lesssim x_{\mathrm{esc}}$, beyond which photons leave the system and $J(x)$ tends towards zero \citep[over the intervals $\pm \lbrack x_{\mathrm{esc}},\,2 x_{\mathrm{esc}}\rbrack$,][]{Adams1972}. Taking this into account only changes the pre-factors by order unity.}). However, a given photon will scatter $\sim x^2$ times at the frequency $\sim x$.
Consequently, $\sim x^2$ scattering events are not \textit{into} a frequency interval which allows for escape, and thus the probability to scatter \textit{into} $[x\pm \dd x / 2]$ is $\sim \phi(x)/x^2 \dd x$. This implies a cumulative escape probability
\begin{equation}
P_{\mrm{esc}} = 2 \int\limits_{x_{\mrm{esc}}}^{\infty}\dd x \frac{\phi(x)}{x^2} = \frac{2 a_v}{3 \pi x_{\mrm{esc}}^3}\;,
\end{equation}
where in the last equality we have used the wing-approximation for $\phi(x)$. The number of scatterings required to escape is $1/P_{\mrm{esc}}$ and plugging in $x_{\mrm{esc}}$ from Eq.~\eqref{eq:xesc_adams} one obtains
\begin{equation}
N_{\mrm{sct}}^{\mrm{esc}} \approx 4.6 \tau_0 \approx 2.7\times 10^6 \left(\frac{N_\HI}{10^{19}\cm^{-2}}\right) \left(\frac{T}{10^4\,\mathrm{K}} \right)^{-1/2}\;.
\label{eq:N_sct_esc}
\end{equation}
This relation differs only by a factor of a few with the exact solution of \citet{1973MNRAS.162...43H} which has been backed up by numerical results \citep[e.g.,][]{1979ApJ...233..649B,2006ApJ...649...14D}.

In summary, \citet{Adams1972} found that a typical \Lya photon leaving
an optically-thick slab scatters a large number of times
essentially in-place (Eq.~\eqref{eq:N_sct_esc}), until reaching the
frequency $x_{\mrm{esc}}$ (Eq.~\eqref{eq:xesc_adams}), after which it
escapes undergoing $N_{\mrm{sct}}^{\mrm{exc}}\sim x^{2}_{\mrm{esc}}$
scattering interactions in the wing of the line.

\subsection{Radiative transfer in clumpy medium}
\label{sec:radi-transf-clumpy}
Resonant line transfer in a clumpy medium has fundamentally different
behavior than in a homogeneous medium, because much of the distance
can be traversed in the optically-thin medium between the clumps.  As
we discussed in the previous section, due to its highly variable
interaction cross section, \Lya escapes through `excursion' in regimes
where the mean free path at the initial frequency is short. In a
multiphase medium, however, a significant fraction of the volume may
have no neutral hydrogen at all. The gas opacity thus varies
strongly with position, even at line center.  This opens up an
alternate escape route in which \Lya photons ``solve the maze'' by
scattering into the optically thin medium between clouds.  This
possibility is essential to consider, since astrophysical systems such
as the ISM and CGM are thought to have a multiphase nature
\citep[e.g.,][]{McKee1977}.

\subsubsection{Model parameters}
In this section, we describe the expected propagation of \Lya photon in a clumpy medium, which we model using spherical clumps of radius $r_\cl$ and \HI number density $n_{\HI,\cl}$ placed in an otherwise empty, semi-infinite slab of height $2B$. 
In what follows, we will use the clump column density $N_{\HI,\cl}=r_\cl n_{\HI,\cl}$ and optical depth $\tau_\cl(x, T) = N_{\HI,\cl}\sigma_\HI(x, T)$ as convenient notation. 
The most important parameter of a clumpy medium is the covering factor \fc which describes the average number of clumps per orthogonal sightline between the midplane and the surface of the slab. These sight-lines will intercept a column density of $N_{\rm HI, total} = \frac{4}{3} \fc N_{\rm HI, cl}$ where the factor $4/3$ is due to the spherical geometry of the clumps\footnote{The mean path length through a sphere of radius $r$ is $\mathrm{Volume} / \mathrm{Area} = 4/3 \pi r^3 / (\pi r^2) = 4/3 r$.}.

\begin{figure}
  \centering
  \includegraphics[width=.95\linewidth]{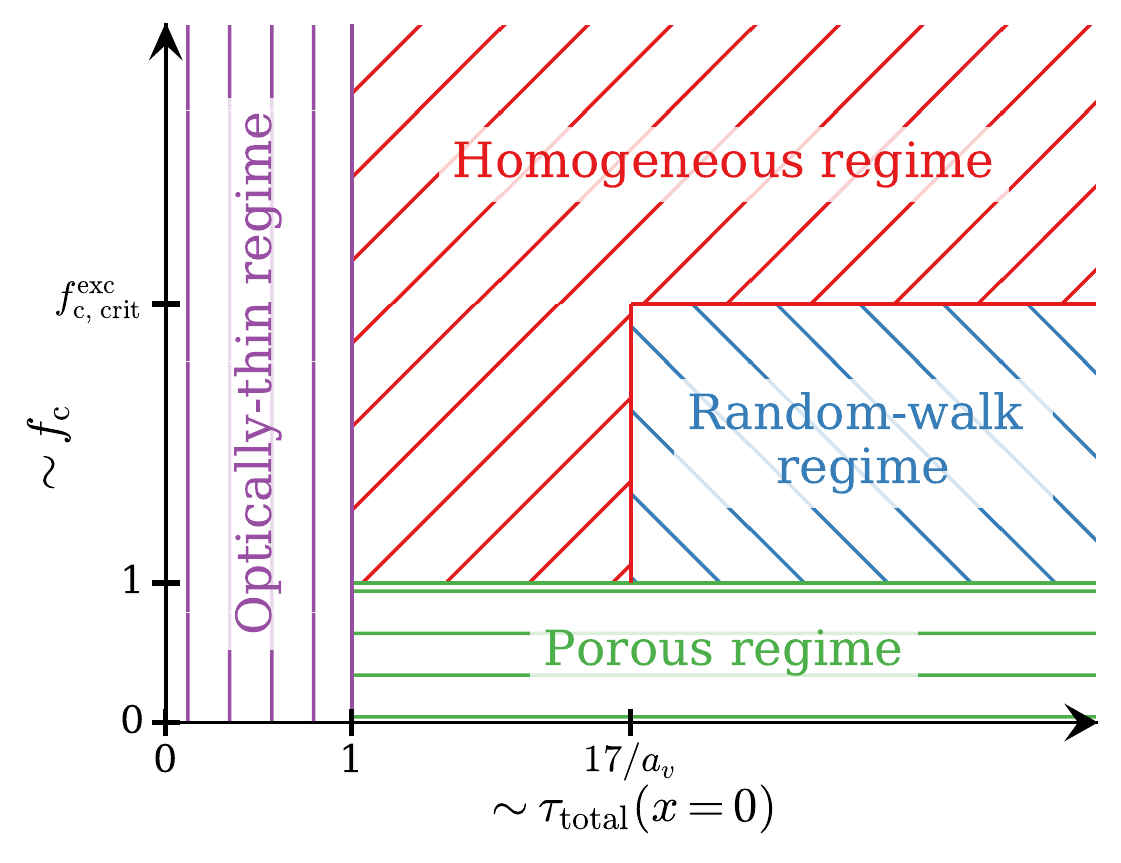}
  \caption{Sketch of radiative transfer regimes in a static, clumpy medium discussed in \S~\ref{sec:radi-transf-clumpy}. The $x$-axis shows the total optical depth at line center and $y$-axis the covering factor \fc.}
  \label{fig:sketch_regimes}
\end{figure}

\subsubsection{Escape regimes}
\label{sec:escape-regimes}
In a static, clumpy medium several regimes are possible for the escape
of a \Lya photon.
We introduced some of them in \citet{clumps1}, but will describe each
regime below. Furthermore, we will sketch \textit{(i)} under which
circumstances each regime is active, \textit{(ii)} on average, how
many clumps a photon encounters $N_{\cl}$, and \textit{(iii)} which emergent spectrum can be expected. Additionally, Fig.~\ref{fig:sketch_regimes} provides a visual overview of the regimes, and a similar overview for a non-static setup is given in Appendix~\ref{sec:regimes_moving}.
\begin{itemize}[itemsep=10pt]
\item \textit{Porous regime.} If a substantial number of sight-lines do not intercept any clumps, many photons will not scatter and simply escape at their intrinsic frequency. 
The fraction of sight-lines without any clumps can be estimated
assuming the clumps are Poisson distributed with mean $\fc$ yielding
$\exp(-\fc)$ \citep[cf][]{Gronke2016a,DijkstraLyaLyC2016}. This area of the
parameter space has been explored in previous work
\citep[][]{Hansen2005,Laursen2012,Gronke2016a} and is of interest as
the empty sight-lines allow for ionizing photon escape
\citep{2015A&A...578A...7V,DijkstraLyaLyC2016} and might allow for directionally dependent
photon escape \citep{Gronke2014a}.  This is the regime suggested by
cosmological simulations of the CGM \citep[e.g.,][]{2015MNRAS.449..987F,2016MNRAS.458.1164L}, though we note
that may be a consequence of their limited resolution, which strongly
limits the number of clumps to be no more than $\sim$ a few.
\item \textit{Random walk regime.} If the clumps are optically thick
  to the photons, i.e., $\tau_{\cl}\gtrsim 1$, the photons are
  expected to scatter at every clump encounter. When $\tau_{\cl}\gg 1$
  the photon scatters close to the surface of each clump and
  effectively random-walks between the clumps, rather than through
  them.  This regime has been studied by \citet{Neufeld1991} and by \citet{Hansen2005}. 
In this `random walk regime' the number of clumps a photon intercepts scales as $\propto \fc^2$. This is because after $N_{\cl}$ interactions a photon has travelled on average a distance $\sqrt{N_{\cl}} B / \fc$ away from the midplane. Thus, to escape this distance has to be $\sim B$ which yields $N_{\cl} = \fc^2$. \citet{Hansen2005} found the scaling to be
\begin{equation}
N_{\cl} \sim \fc^2 + \fc,
\label{eq:N_cl_generic}
\end{equation}
with pre-factors of order unity which depend somewhat on the geometry
(see \citealt{Hansen2005} for details).
\item \textit{Optically-thin regime.} If the clumps are, on the other hand, optically thin to the intrinsic radiation ($\tau_{0,\cl}\lesssim 1$), not every cloud interception will necessarily cause the photon to scatter. In particular, the probability of a scattering event to happen in this case is $1 - \exp(-\tau_{0,\cl}) \approx \tau_{0,\cl}$. This implies that in order to describe the expected scaling in this regime, we can replace in the above considerations $\fc$ by $\tau_{0,\cl}\fc$. As in this regime $\tau_{0,\cl}\lesssim 1$, this corresponds to effectively decrease the pre-factors in Eq.~\eqref{eq:N_cl_generic}. Specifically, if \textit{all clumps} in a given sightline are optically thin the intrinsic photons (i.e., at line center), that is, $\tau_{0,\mrm{total}}\equiv 4/3 \fc \tau_{0,\cl} \lesssim 1$ most \Lya photons will not interact before escaping. This means they will simply stream through all the clumps (leading to $N_\cl \propto \fc$) keeping their intrinsic frequency (i.e., a peak frequency of $x_{\mrm{p}}\sim 0$). We call this `optically-thin' regime.
\item \textit{Homogeneous regime.} Since $\tau_{\cl}$ depends strongly
  on the frequency of the photon which changes throughout the
  scattering process, \Lya might also escape from clumpy media in a
  frequency excursion as discussed in the homogeneous slab. In
  particular, during the course of the $\sim\fc^2$ scatterings needed
  to random-walk through the clumpy medium, the photon may scatter far
  enough into the wing of the line to escape the medium in a single excursion as described in \S~\ref{sec:radi-transf-homog}. If this happens, most clumps become optically thin for the photon and  one can generalize the argument made above when describing the `optically thin regime' by replacing $\tau_{\mrm{0,cl}}$ by $\tau_{\cl}(x)$. 
Since this possibility becomes increasingly likely with every
scattering event, we anticipate that above some critical value of
$\fc$, clumpy media behave like homogeneous slabs in the sense that
photos escape via frequency excursion.
\end{itemize}

Conclusively, the four regimes are different in their preferential
escape route of the photons which depends in the static case on the
covering factor \fc and on the optical depth of individual clumps. Each escape route implies that the photon experiences a clumpy medium differently which leaves a clear imprint on the emergent spectrum. One way to characterize these regimes is: the `optically thin' and `porous' regimes represent escape without significant interaction, while the `random walk' and `homogeneous' regimes represent escape primarily via spatial or frequency diffusion respectively.
Table~\ref{tab:regimes} provides a brief summary of the different regimes.

\subsubsection{Division between the regimes}
\label{sec:divis-betw-regim}
In the last section we introduced the four different routes for a \Lya
photon to escape from a clumpy medium. We also briefly discussed the
physical conditions under which each escape route is favored.  In this
section, we quantify these boundaries more precisely.

We denote the boundary between the homogeneous and the other regimes
with \fccrit, and specifically to the random walk regime with
$\fccrit^{\mrm{exc}}$. Physically, this value characterizes the
critical number of clumps per sightline when a excursion-like escape
becomes faster than a random-walk diffusion. In order to find this
boundary let us follow this argument and compute the criteria for  when it is possible for the photons to stream through the clumps\footnote{This derivation of \fccrit in a static setup is complementary to the one presented in \citet{clumps1} where we used a time-scale argument.}. 
As stated above, the characteristic escape frequency is given by Eq.~\eqref{eq:xesc_adams}, and in a clumpy medium the total line center optical depth is $\tau_0 = 4/3 f_c N_{\HI,\cl} \sigma_0$. The transition occurs when photons can stream through individual clumps, that is when $4/3\tau_{\cl}(x_{\mathrm{esc}}) = 1$. Using the wing approximation for the \Lya cross section, this yields
\begin{equation}
\fccrit^{\mathrm{exc}} = \frac{2 \sqrt{a_v \tau_{0, \cl}}}{3
  \pi^{1/4}}
\approx \left(\frac{N_{\HI,\cl}}{10^{17}\,\text{cm}^{-2}}\right)^{1/2}\left(\frac{T}{10^{4}\,\text{K}}\right)^{-1}\;.
\label{eq:fccrit_exc}
\end{equation}
For optically thinner medium an escape through excursion is not possible as a frequency shift into the wings of the lines will lead to immediate escape. Specifically, this happens if the wings become optically thin, i.e., if $\sqrt{3}\tau(x_*) < 1$ which translates to $\sqrt{3}\tau_0 a_v < \sqrt{\pi} x_*^2 \approx 18.78$ (where we included factors of $\sqrt{3}$ due to the rectangular geometry). This transition happens for an homogeneous medium as well as a clumpy medium and sets a lower limit to \fccrit. However, if the individual clumps possess an optical depth at line center of $\tau_{0,\cl} \lesssim 1$, not every clump encounter leads necessarily to a scattering, and, thus introduces the additional factor of $1-\e^{-\tau_{0,\cl}}$ (as described for the \textit{optically-thin} regime in \S~\ref{sec:divis-betw-regim}).

In conclusion, we expect the transition to the homogeneous regime to occur if
\begin{equation}
\fccrit =  \begin{dcases}
\fccrit^{\mrm{exc}} = \frac{2 \sqrt{a_v \tau_{0, \cl}}}{3 \pi^{1/4}}  & \text{ for } \sqrt{3} a_v \tau_0 > \sqrt{\pi} x_*^2\\
\frac{\fccrit^{\mrm{sf}}}{1-\e^{-\tau_{0,\cl}}}  & \text{ otherwise.}
\end{dcases}
\label{eq:fccrit} 
\end{equation} 
where $\fccrit^{\mrm{sf}} =2 x_*/ 3^{5/4} \approx 1.65$ is due to continuity of \fccrit at $\sqrt{3}a_v\tau_0 = \sqrt{\pi} x_*^2$, that is, at the transition from excursion to single-flight escape.

\subsubsection{Non-static case}
\label{sec:non-static-case}
Since the \Lya cross section depends sensitively on the frequency $x$, clump motions can dramatically influence radiative transfer.
If a clump moves with a velocity
$\gtrsim{x_*}v_{\text{th}}\sim{50}$\,km/s,  a single clump interaction
will put the photon far enough into the wing of the line to 
allow the photon to escape directly in a single excursion. This
possibility is important to consider because random velocities
$v\sim{100}$\,km/s may be typical for the CGM in galaxies and in
galactic winds, and velocities $v\sim{1000}$\,km/s may be typical in
the regions around black holes.
This means that, for large $\fc$ the medium behaves as a slab with an increased temperature of 
\begin{equation}
T_{\rm eff} = T + \frac{\sigma_{\cl}^2 m_H}{2 k_B}
\label{eq:Teff}
\end{equation}
where $\sigma_\cl$ is the 1$D$ velocity dispersion of the clumps.

For a lower number of clumps the overall velocity distribution is not well-sampled which leads to sight-lines with no clumps in the core of the line. In this case the photons escape without any clump interaction.
We can estimate this to happen if the mean separation of two clumps in velocity space becomes larger than the velocity range over which a clump can provide $\tau_{\cl}\gtrsim 1$. For a Gaussian velocity distribution with variance of $\sigma_\cl^2$ the average separation is approximately given by $\sigma_\cl / (\alpha\fc)$ where $\alpha$ is the fraction of clumps within core of the velocity distribution, i.e., in our case $\alpha\approx 0.68$. Consequently, the transition to the homogeneous regime for randomly moving clumps occurs at $4/3\tau_\cl(\sigma_\cl / (\alpha \fc v_{\mrm{th}})) = 1$ which -- using the core approximation and including geometrical factors -- can be written as a critical covering factor for the randomly moving case
\begin{equation}
  \label{eq:fccrit_moving}
  \fccrit = 
  \begin{dcases}
\frac{\sigma_\cl}{\alpha v_{\rm th} \sqrt{\ln(4/3 \tau_{0,\cl})}} & \text{if }\fc > 1/\alpha\\
\frac{1}{\alpha} & \text{otherwise.}
  \end{dcases}
\end{equation}
Here, the lower boundary of $1/\alpha$ results simply from the requirement that at least one clump within the core of the velocity distribution function is necessary in order to sample the core of the velocity distribution. We expect for larger covering factors the system to behave as a homogeneous slab of temperature $T_{\mathrm{eff}}$. See also Appendix~\ref{sec:regimes_moving} for more details about the expected behavior in the case of uncorrelated  clump motion.\\

In the case of clumps with a systematic velocity structure (for instance, outflowing clumps), the above requirement of a `well-sampled' velocity field is fulfilled if the adjacent clump is optically thick to the \Lya photon, i.e., if $4/3\tau_\cl(x_{\rm next}) \gtrsim 1$ where $x_{\mathrm{next}}$ depends on the exact velocity profile. For a linearly scaled (Hubble-like) outflow from $0$ at 
midplane to $|v_{\mathrm{max}}|$ at the boundaries of the slab, we have $x_{\rm next} = v_{\rm max} / (\fc v_{\rm th})$. 
In addition, a photon might be artificially forced into the wing of the line if $x_{\rm next} > x_*$ due to the sampling of the velocity field. This does not occur in a homogeneous medium, and thus, for a Hubble-like outflow the criterion to be fulfilled in order the be in the homogeneous regime is
\begin{equation}
  \label{eq:fccrit_outflow}
  \fccrit = 
  \begin{dcases}
\frac{\sqrt{\pi} v_{\mathrm{max}}^2}{a_v N_{\HI, \mathrm{total}} v_{\mathrm{th}}^2 \sigma_0} & \text{if }v_{\mathrm{max}} > \hat v_{\mathrm{max}}\\
x_* v_{\mathrm{th}}/v_{\mathrm{max}} & \text{otherwise.}
  \end{dcases}
\end{equation}
where $\hat v_{\mathrm{max}} = v_{\mathrm{th}}\left(a_v N_{\HI, \mathrm{total}} x_* \sigma_0/\sqrt{\pi}\right)^{1/3}$.


\section{Numerical Method}
\label{sec:method}

\subsection{\Lya radiative transfer}
\label{sec:lya-radi-transf}
Due to the complexity of the resonant line transfer, Monte-Carlo radiative transfer codes are commonly in use \citep[e.g.,][]{Auer1968,Ahn2001a,Zheng2002}. This algorithm works by following individual photon packages in a stochastic manner through real- and frequency-space until their escape.
In this work, we use the code \texttt{tlac} which has been used and described previously, e.g., in \citet{Gronke2014a}. In particular, we make use of \texttt{tlac}'s features \textit{(i)} to handle embedded spherical grids within a Cartesian grid,  and \textit{(ii)} employ a dynamical core-skipping scheme \citep[as described in][]{Smith2014,Gronke2016a}. We also turned off the dynamical core-skipping for a few models and checked that the emergent spectra are identical.
We run most setups using $\sim 10^4$ photon packages but use occasionally more to obtain a higher-resolution spectrum.


\begin{table}
  \centering
  \caption{Overview of the model parameters}

  \newcommand*{\myspace}{\addlinespace[0.5em]}

  \begin{tabularx}{\linewidth}{lLl}
    \toprule
    \heading{Symbol} 
    & \heading{Description} 
    & \heading{Considered values}
    \\
    \midrule

    $B$ 
    & Half-height of slab 
    & $50\,$pc
    \\ \myspace

    $r_{\cl}$ 
    & Clump radius 
    & $\{10^{-3},\,10^{-2}\}\,$pc
    \\ \myspace

    $\fc$\tablefootmark{a} 
    & Covering factor 
    & $[0,\,2000]$
    \\ \myspace

    $T$ 
    & Temperature 
    & $10^4\,$K
    \\ \myspace

    $N_{\rm HI, cl}$\tablefootmark{b} 
    & Clump \HI column density
    & $[10^{12},\,10^{22}]\cm^{-2}$
    \\ \myspace

    $\sigma_{\cl}$ 
    & Clumps' velocities standard deviation 
    & $[0,\,500]\kms$
    \\ \myspace

    $\tau_{\rm d, cl}$\tablefootmark{b} 
    & Clump (absorbing) dust optical depth  
    & $[10^{-4},\,1]$
    \\ \myspace

    $v_{\rm max}$ 
    & Maximum clump outflow velocity
    & $[0,\,5000]\kms$
    \\ \myspace

    \cmidrule(l{.7cm}r{.7cm}){1-3}
    \myspace

    $F_V$ 
    & Volume filling factor 
    & $[0,\,0.6]$
    \\ \myspace

    $N_{\rm HI, total}$\tablefootmark{a}
    & \HI column density of slab 
    & $[10^{11},\,10^{26}]\cm^{-2}$
    \\ \myspace

    $n_{\rm HI, cl}$ 
    & \HI number density in clumps 
    & $[10^{-5},\,10^{6}]\cm^{-3}$
    \\ \myspace

    $\tau_{0, {\cl}}$\tablefootmark{b} 
    & Line center optical depth of clump 
    & $[0.05,\,10^{9}]$
    \\ \myspace

    $\tau_{\mrm{0,\,total}}$\tablefootmark{a} 
    & Total line center optical depth 
    & $[10^{-3},\,10^{13}]$ 
    \\ \myspace
    
    $\tau_{\mrm{d,\,total}}$\tablefootmark{a}
    & Total (absorbing) dust optical depth 
    & $[0,\,10^3]$ 
    \\

    \bottomrule
  \end{tabularx}

  \tablefoot{Above and below the horizontal line are the free and dependent parameters, respectively.\\
    \tablefoottext{a}{From emitting plane to boundary of the slab, i.e., per $B$.}\\
    \tablefoottext{b}{From center to boundary of a clump, i.e., per $r_{\cl}$.}\\
  }

  \label{tab:params}
\end{table}

\subsection{Model parameters}
\label{sec:params}
Analogous to \S~\ref{sec:radi-transf-clumpy}, our setup consists of a slab with half-height $B$ in which we distribute spherical clumps with radius $r_{\cl}$ randomly in the box until a fraction of the total volume $F_V$ is filled. This means the number density of clumps is $n_\cl = F_V / (4/3 \pi r_\cl^3)$ where $r_\cl$ is the clump radius. 
The connection between the volume filling factor $F_V$ and the previously introduced covering factor \fc (which describes the average number of clumps a line orthogonal to the slab intercepts between the midplane and the boundary of the box) is given by the 
integration along the finite axis of the slab, that is,
\begin{equation}
\fc = \int\limits_0^B \dd r\; \pi n_\cl r_\cl^2 = \frac{3 F_V B}{4 r_{\cl}}\;.
\end{equation}

The clumps are filled with neutral hydrogen with a number density of $n_{\rm HI, cl}$ and temperature $T$, leading to a column density between the center of the clumps to their outskirts of $N_{\rm HI, cl} = r_{\cl} n_{\rm HI, cl}$. As described in \S~\ref{sec:radi-transf-clumpy}, this means on average the shortest path between the midplane and the boundary of the box will intercept a column density of
\begin{equation}
N_{\rm HI, total} = \frac{4}{3} \fc N_{\rm HI, cl} = F_V B n_{\rm HI, cl}\;.
\end{equation}

In general, we consider three cases: the static case with no motion, the randomly moving case, and an outflowing case.
In the randomly moving case, we assign each clump a random velocity by drawing each  component from a Gaussian with standard deviation $\sigma_{\cl}$. This represents a ``white noise'' spectrum, with velocity differences which are statistically equally probable on all spatial scales. 
For the outflow, we choose a simple linear velocity scaling from $0\kms$ to $v_{\rm max}$ at the midplane and boundary of the slab, respectively.
We will investigate models with correlated turbulence, and different velocity profiles in future work.

Furthermore, we study two different emission sites for the \Lya photons. First, simply the midplane of the box, and secondly randomly chosen emission within the clumps. While the former is useful in order to study merely the radiative transfer processes through the clumpy medium from an external source such as a star-forming region, the latter case represents a physically motivated scenario in which \Lya are produced via recombination events within the clumps. Both scenarios might be responsible, e.g., for the \Lya halos found around galaxies \citep[e.g.][]{Dijksta2012MNRAS.424.1672D,Lluis2016ApJ...822...84M}.


\begin{figure}
  \centering
  \includegraphics[width=.95\linewidth]{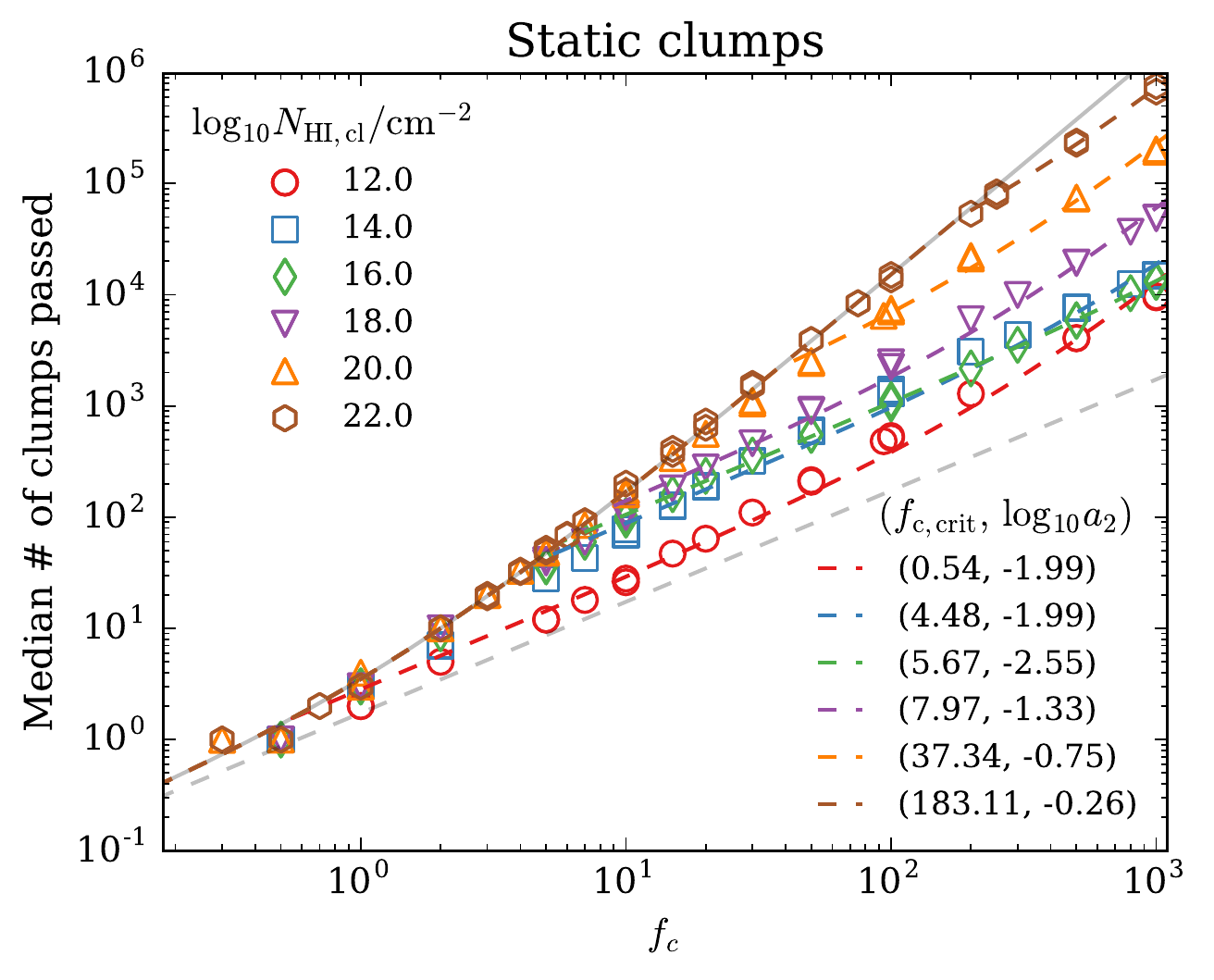}
  \caption{Number of clumps passed versus covering factor $\fc$ for different clump column densities. The dashed lines show fits of Eq.~\eqref{eq:fit_N_cl} to the data points and the grey solid [dashed] line shows the limit with $f_{\rm c, crit} > 10^3$ [$N_{\HI,\cl}\rightarrow 0$].}
  \label{fig:nclumps_static}
\end{figure}

\begin{figure}
  \centering
  \includegraphics[width=.95\linewidth]{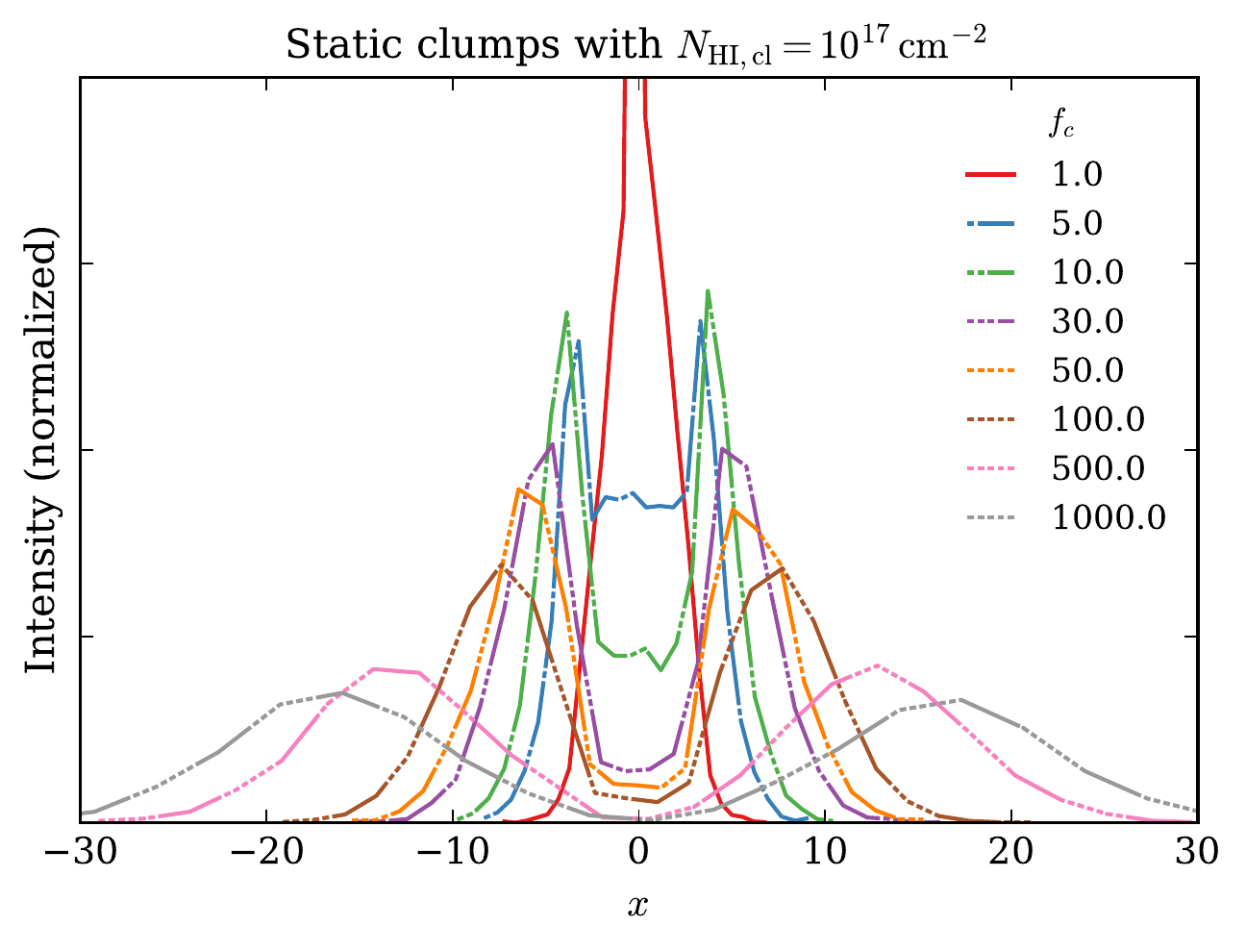}
  \caption{\Lya spectra for a constant clump column density $N_{\rm HI,cl}=10^{17}\cm^{-2}$ and various values of $\fc$ (increasing $\fc$ corresponds to an increased spectral width).}
  \label{fig:spectra_fc}
\end{figure}

\section{Numerical Results}
\label{sec:results}
In this section we present the results from our numerical radiative transfer simulations. In particular, we focus on three quantities, namely the number of clumps encountered by the photons $N_\cl$, and the emergent \Lya spectra. $N_\cl$ is a useful diagnostic, since we expect $N_\cl \sim \fc^2$ for escape via random walk in position space, and $N_\cl\sim\fc$ for escape via excursion and single-flight as described in the previous section.

The section is approximately ordered by ascending complexity. In \S~\ref{sec:static-case}, we discuss the static case, in \S~\ref{sec:random-motion} and \S~\ref{sec:outflows} we introduce random clump motions and outflows, respectively. Moreover, in \S~\ref{sec:clump-emission} we change the emission site of the photons to be inside the clumps which resembles a case of fluorescent emission. Finally, we study the effect of dust inside the clumps on the \Lya escape in a clumpy medium (\S~\ref{sec:dust-within-clumps}) which we quantify through the \Lya escape fraction. 

\begin{figure}
  \centering \includegraphics[width=.95\linewidth]{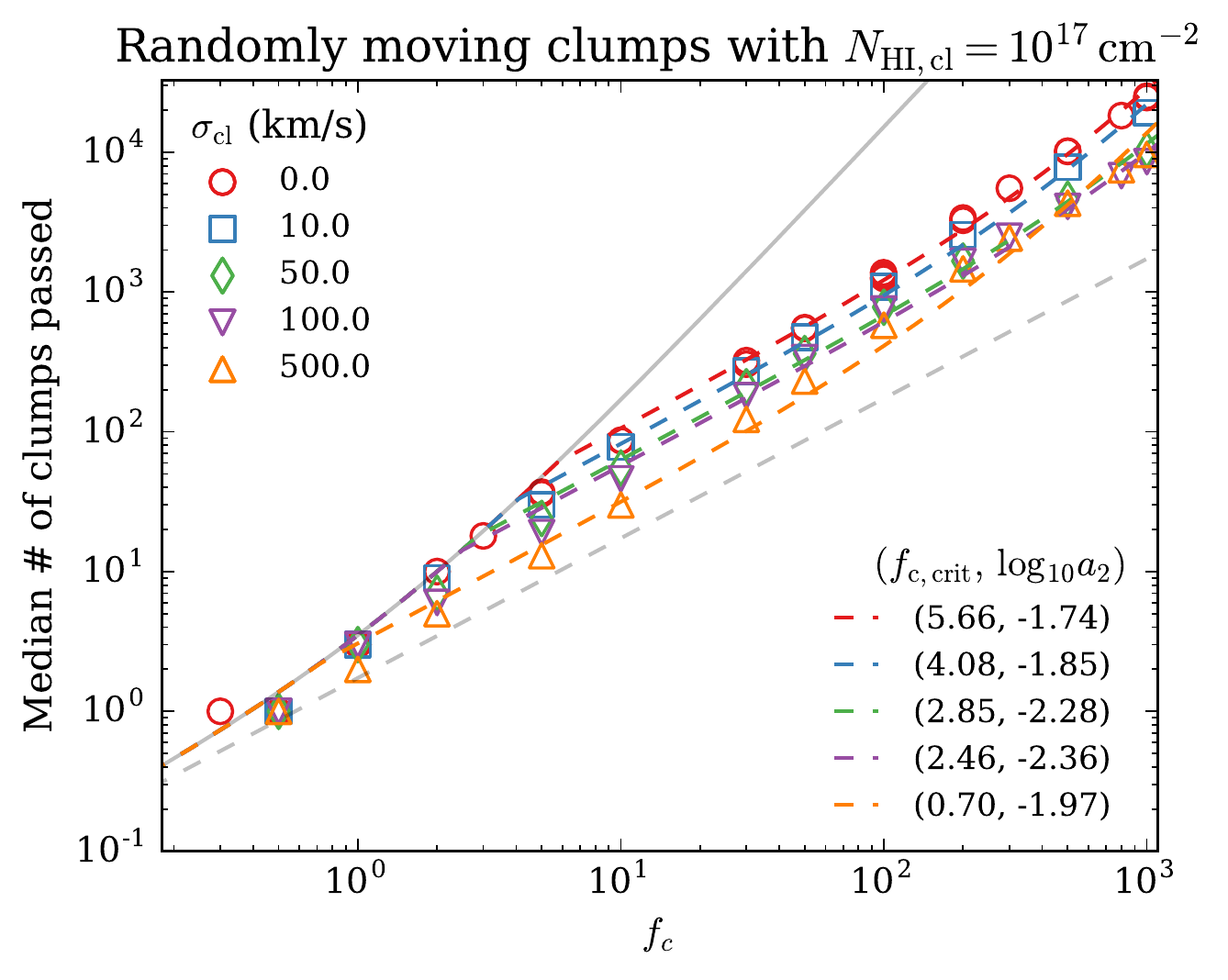}
  \caption{Number of clumps passed versus $\fc$ for clumps with $N_{\rm HI, cl}=10^{17}\cm^{-2}$ and uncorrelated, random motion with various $\sigma_{\cl}$. The dashed lines show fits of Eq.~\eqref{eq:fit_N_cl} to the data points and the grey solid line shows the limit with $f_{\rm c, crit} > 10^3$.}
  \label{fig:nclumps_moving}
\end{figure}

\begin{figure}
  \centering
  \includegraphics[width=.95\linewidth]{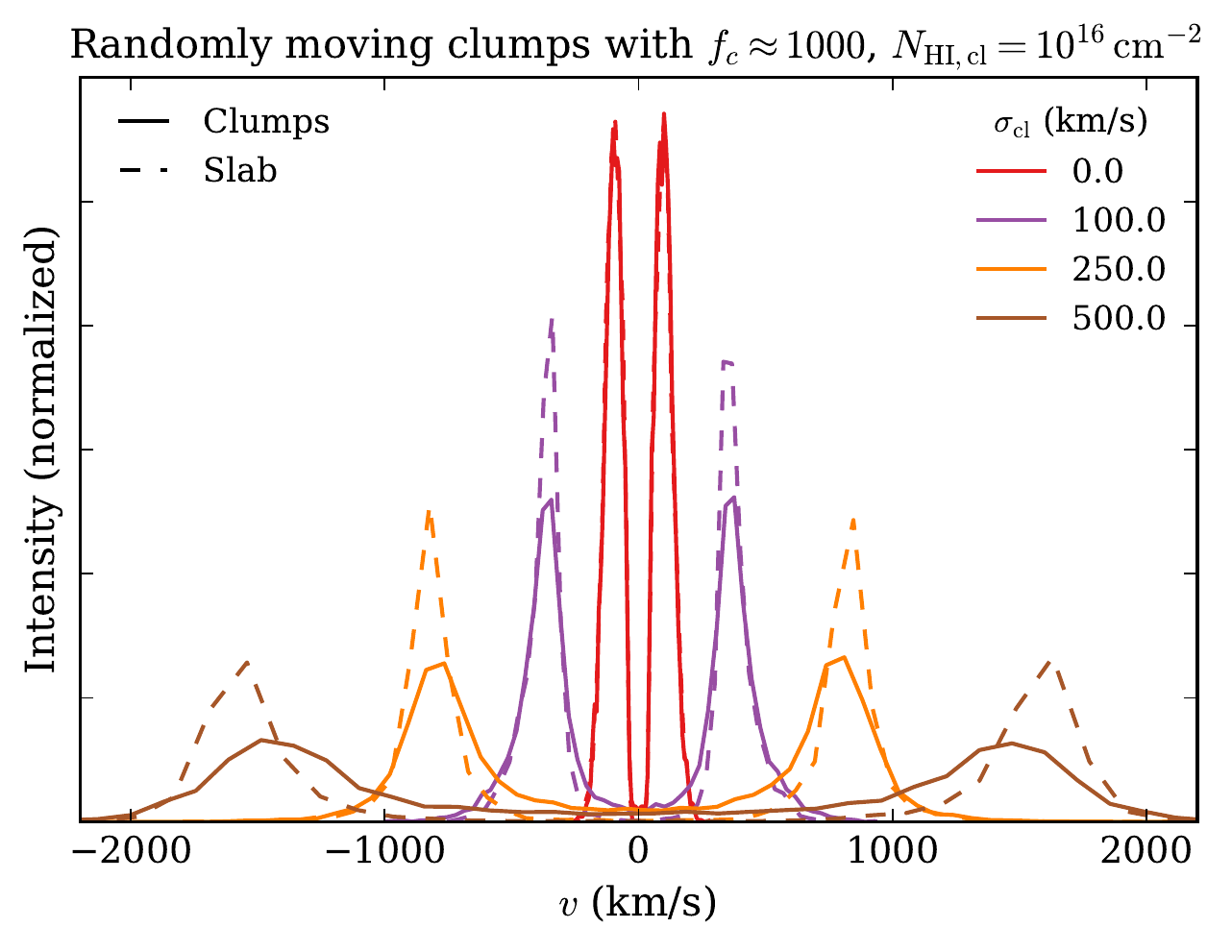}
  \caption{The \textit{solid lines} show the \Lya spectra for a constant clump column density $N_{\rm HI,cl}=10^{16}\cm^{-2}$, and covering factor $\fc\approx 1000$. The \textit{dashed lines} show as comparison the spectra obtained from slabs with $T_{\rm eff}(\sigma_{\cl})$.}
  \label{fig:spectra_vs_slabs}
\end{figure}

\subsection{Static case}
\label{sec:static-case}

Fig.~\ref{fig:nclumps_static} shows the number of clumps a \Lya photon passed through before escaping the box versus the covering factor $\fc$ which we vary over $\sim 3$ orders of magnitude. Each symbol and color represents different values of $N_{\rm HI, cl}$ and, thus, different clump optical depths at line center $\tau_{cl, 0}$ which we vary from $\sim 0.06$ (optically thin) to $\sim 6\times 10^8$ (optically thick). Note, that we also ran each combination of $(N_{\rm HI, cl},\,\fc)$ with two different cloud radii $r_{\cl}=\{10^{-2},\,10^{-3}\}\,$pc to confirm that this parameter is not important \citep{Hansen2005}.

The dashed lines in the corresponding color show curves following
\begin{equation}
  N_{\cl} =  \begin{cases}  
a_1 \fc^2 + b_1 \fc & \text{ for } \fc < \fccrit\\
a_2 \fc^2 + b_2 \fc & \text{ for } \fc \ge \fccrit
\end{cases}\;.
\label{eq:fit_N_cl}
\end{equation}
We fit the data points for $N_{\rm HI, cl}=10^{22}\cm^{-2}$ for $\fc \le 100$ to determine $(a_1\,b_1) = (3/2,\,2)$ (the best fit values are $(1.51,\,1.90)$ which -- given the uncertainty -- we rounded to the nearest convenient fraction for simplicity).
These coefficients represent geometrical factors in the surface scattering regime (where clouds are optically thick), and thus independent of $N_{\rm HI,cl}$. We have directly verified this numerically.
These values are then fixed for all $N_{\rm HI, cl}$ in order to fit each $N_{\rm HI, cl}$-curve for $(\fccrit,\,a_2)$ while $b_2$ is fixed by requiring continuity at \fccrit. Fig.~\ref{fig:nclumps_static} shows the resulting fits as well as the obtained values for $(\fccrit,\,a_2)$. The break in the scaling relation at \fccrit is clearly visible (for visual aid, Fig.~\ref{fig:nclumps_static} also shows the $\fc = a_1\fc^2 + b_1\fc$ curve from which the scaling departures for $\fc > \fccrit$).

We find that the obtained \fccrit for high column densities ($N_{\rm HI, cl}\gtrsim 10^{20}\cm^{-2}$) matches the prediction from \S~\ref{sec:divis-betw-regim} reasonably well, this breaks down for lower optical depths.
Also, for $N_{\rm HI, cl}=10^{12}\cm^{-2}$, i.e., when the clumps are always optically thin for \Lya photons, we obtain $a_2\approx \tau_{0, cl}^2 \approx 0.010$ as discussed in Sec.~\ref{sec:analyt-cons-new}. With increasing $N_{\rm HI, cl}$ we find a decreasing $a_2$ to match the data. Thus, we can identify the escape regimes (characterized by the number of clumps encountered) described in the analytic model in Sec.~\ref{sec:analyt-cons-new}.
In summary, Fig.~\ref{fig:nclumps_static} shows that for $\fc < \fccrit$, $N_{\cl} \propto \fc^2$ (as expected for a random walk), while for $\fc > \fccrit$, $N_\cl \propto \fc$ (as expected for escape through excursion).

Fig.~\ref{fig:spectra_fc} shows the corresponding spectra for $N_{\rm HI, cl}=10^{17}\cm^{-2}$. Note, that for this column density we found $\fccrit \sim 2$ which corresponds roughly to the boundary between single and double peaked spectra. In particular, we recover the spectral shape of \citet{Hansen2005} for $\fc \ll \fccrit$ while obtaining wide, double peaked spectra with zero flux at line center for $\fc \gg \fccrit$. This means, that the escape regimes do not only impact the photons' paths but modify the escape frequencies', and hence, leave a clear observational signature on the emergent spectra.

\begin{figure}
  \centering \includegraphics[width=.95\linewidth]{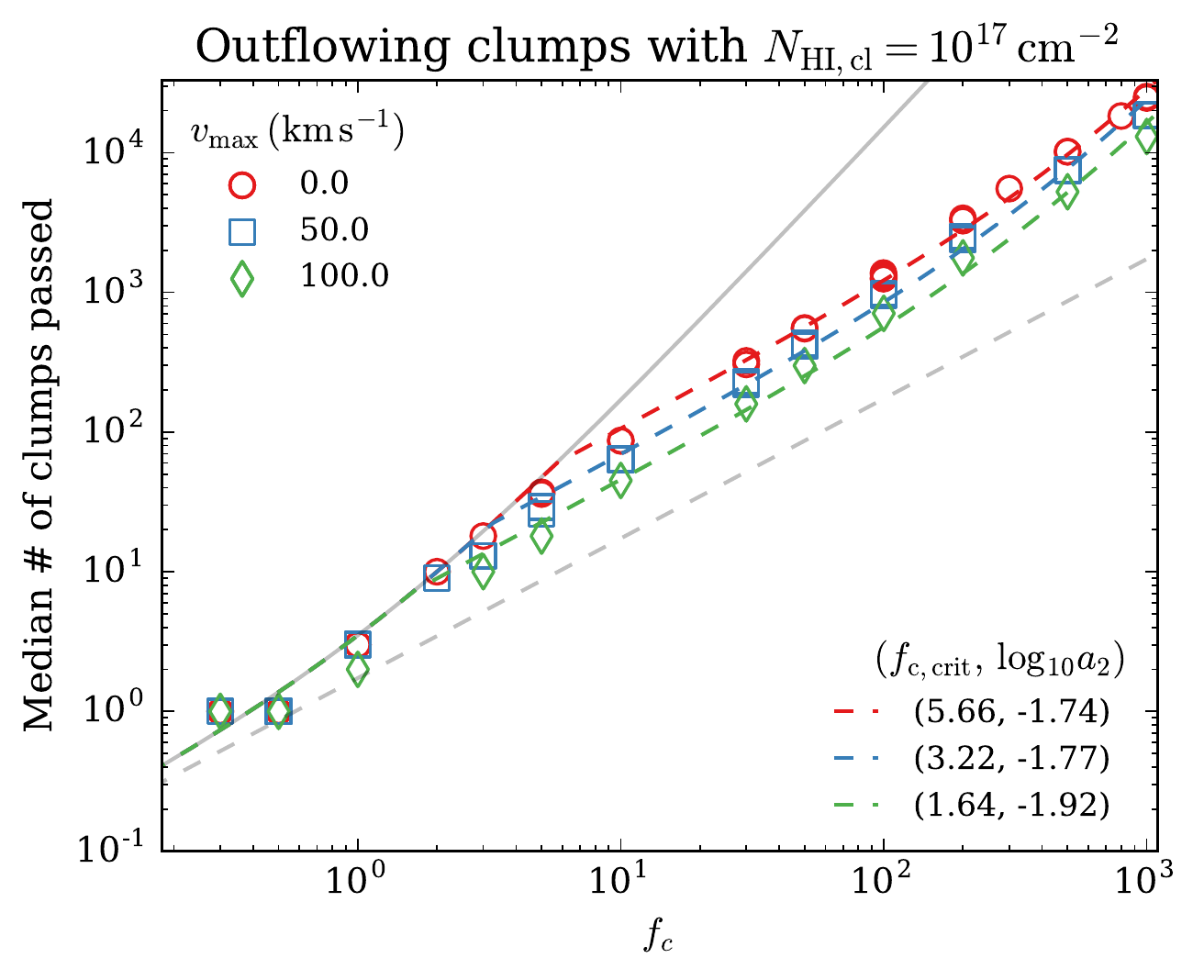}
  \caption{Number of clumps passed versus $\fc$ for clumps with $N_{\rm HI, cl}=10^{17}\cm^{-2}$ and outflowing motions with different maxima $v_{\rm max}$. The dashed lines show fits of Eq.~\eqref{eq:fit_N_cl} to the data points and the grey solid line shows the limit with $f_{\rm c, crit} > 10^3$.}
  \label{fig:nclumps_outflow}
\end{figure}

\begin{figure*}
  \centering
  \includegraphics[width=.95\linewidth]{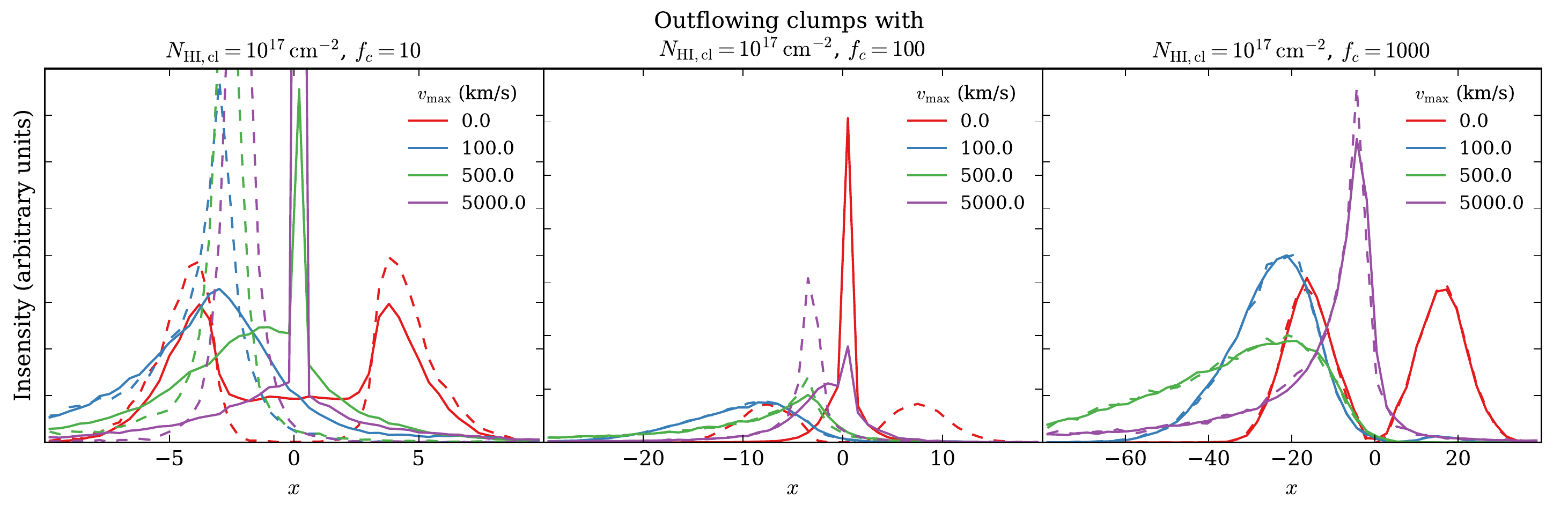}
  \caption{\Lya spectra using a setup of outflowing clumps with linear velocity profile for $N_{\rm HI, cl}=10^{17}\cm^{-2}$ and four different maximal velocities $v_{\rm max}$. The \textit{dashed lines} in corresponding colors show the spectra emergent from a slab with the same column density and velocity structure. Each sub-panel displays a case with different covering factor corresponding to increasing agreement with the homogeneous setup.}
  \label{fig:outflows}
\end{figure*}

\subsection{Random motion}
\label{sec:random-motion}

Fig.~\ref{fig:nclumps_moving} shows the $N_{\cl}-\fc$ scaling relation in the case of random clump motion for a fixed clump's column density of $N_{\rm HI, cl}=10^{17}\cm^{-2}$. 
As conjectured in \S~\ref{sec:non-static-case}, compared to the static case the photons spend less time until escape and thus the number of clumps passed is smaller. This is due to the fact that in the case of fast moving clumps, the photons escape either through `holes' in velocity space (where, $N_\cl \propto \fc$), or via single flight (in which case also $N_\cl \propto \fc$). Departures from that are either due to convergence to the static case (for $\sigma_\cl \rightarrow 0$), or when escape via single flight involves multiple surface scatterings (for $\fc\ll\fccrit$) when the interaction with another clump is non-negligible.

In Fig.~\ref{fig:spectra_vs_slabs} we show the emergent spectra from this setup. In particular, we focus on the case with $\fc\approx 1000$ and $N_{\rm HI, cl}=10^{16}\cm^{-2}$ and four different values of clump velocity dispersion $\sigma_{\cl}$. Also in Fig.~\ref{fig:spectra_vs_slabs} we overlay spectra from homogeneous slabs with an effective temperature $T_{\rm eff}$ (see Eq.~\eqref{eq:Teff}) corresponding to the respective value of $\sigma_{\cl}$. Clearly, the spectra match quite well -- especially the peak separation. However, with increasing $T_{\rm eff}$ the matches become worse, which makes sense, since the wider velocity space is more poorly sampled.

\begin{figure}
  \centering
  \includegraphics[width=.95\linewidth]{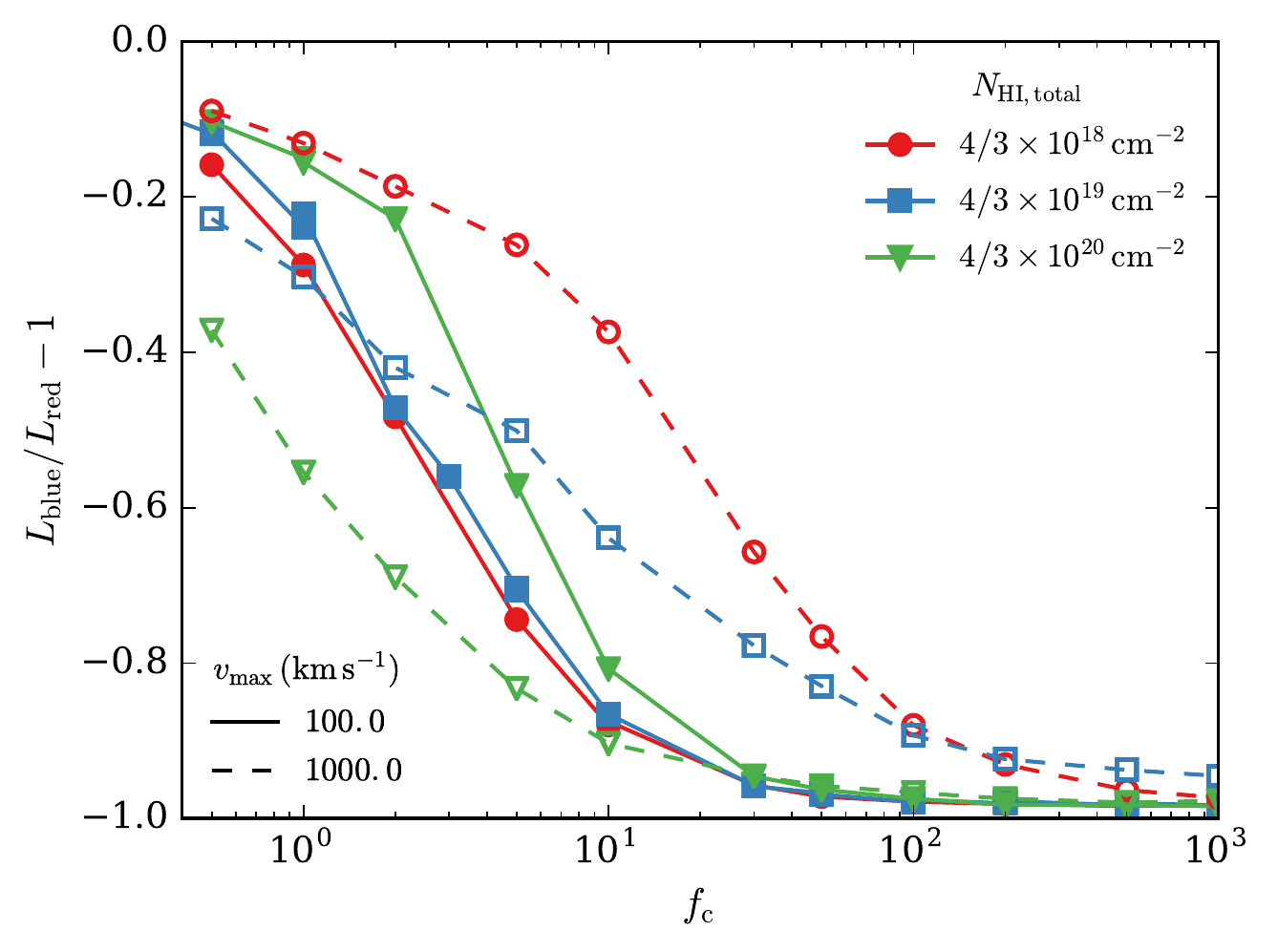}
  \caption{Integrated blue over integrated red flux (minus one) versus covering factor for different combinations of $v_{\rm max}$ and $N_{\HI, \mathrm{total}}$. With increasing \fc the spectra become more redshifted. See \S\ref{sec:outflows} for details}
  \label{fig:outflow_asymmetry_cut}
\end{figure}

\subsection{Outflows}
\label{sec:outflows}
Fig.~\ref{fig:nclumps_outflow} shows the $N_{\cl}-\fc$ relation in the presence of linearly scaled outflows with maximum velocity $v_{\rm max}$ (as described in \S~\ref{sec:params}). We can see that a flattening of the curve still exists which we interpret again as the transition between the `random walk' and `homogeneous regime'.
As expected, with increasing outflow speed  this threshold decreases.

Fig.~\ref{fig:outflows} illustrates the change in spectral shape when introducing outflows. In each subpanel, the solid lines show the emergent spectrum from the clumpy model and the dashed lines in matching color the ones from a homogeneously filled slab with the same total column density and velocity structure.
We focus on the case with constant clump column density $N_{\rm HI, cl}=10^{17}\cm^{-2}$ and show three cases $f_{\rm c}=\{1,\,100,\,1000\}$ which match increasingly well the slab case -- for all four values of $v_{\rm max}$.
This implies that the spectra become more asymmetric as they converge towards the homogeneous limit. The asymmetry develops because the outflow shifts the scattering cross section in the observers reference frame towards the blue. Thus, the optical depth for photons with frequency redward of line-center (e.g., `back-scattered' ones off the far-side of the system) is lowered allowing for easier escape.
We will discuss the result of higher asymmetry with increased number of clumps further in \S~\ref{sec:discussion}.

Fig.~\ref{fig:outflow_asymmetry_cut} shows this increase in asymmetry with greater \fc for fixed outflow velocities of $v_{\mathrm{max}} = \{100,\,1000\}\kms$ and total column densities of $N_{\HI, \mathrm{total}}=4/3\times\{10^{18},\,10^{19},\,10^{20}\}\cm^{-2}$. We characterize the spectral asymmetry by the integrated flux ratio of the blue over the red part of the spectra (minus one), i.e., a value of $-1$ means that all photons escape redward of line center ($x \le 0$) and if this quantity is zero the spectrum is symmetric (around $x = 0$). While the transition from symmetric to dominantly red spectra for $v_{\mathrm{max}}=100\kms$ is nearly independent of the column density at $\fc \sim 10$, this is not the case for $v_{\mathrm{max}}=1000\kms$ where a larger total column density implies a shift at lower \fc. This is due to the dependence of \fccrit on $v_{\mathrm{max}}$ and $N_{\HI, \mathrm{total}}$ described in \S~\ref{sec:non-static-case}. In particular, as seen in Eq.~(\ref{eq:fccrit_outflow}) \fccrit does not depend on the column density if $v_{\mathrm{max}}$ is small.

\begin{figure}
  \centering
  \includegraphics[width=.95\linewidth]{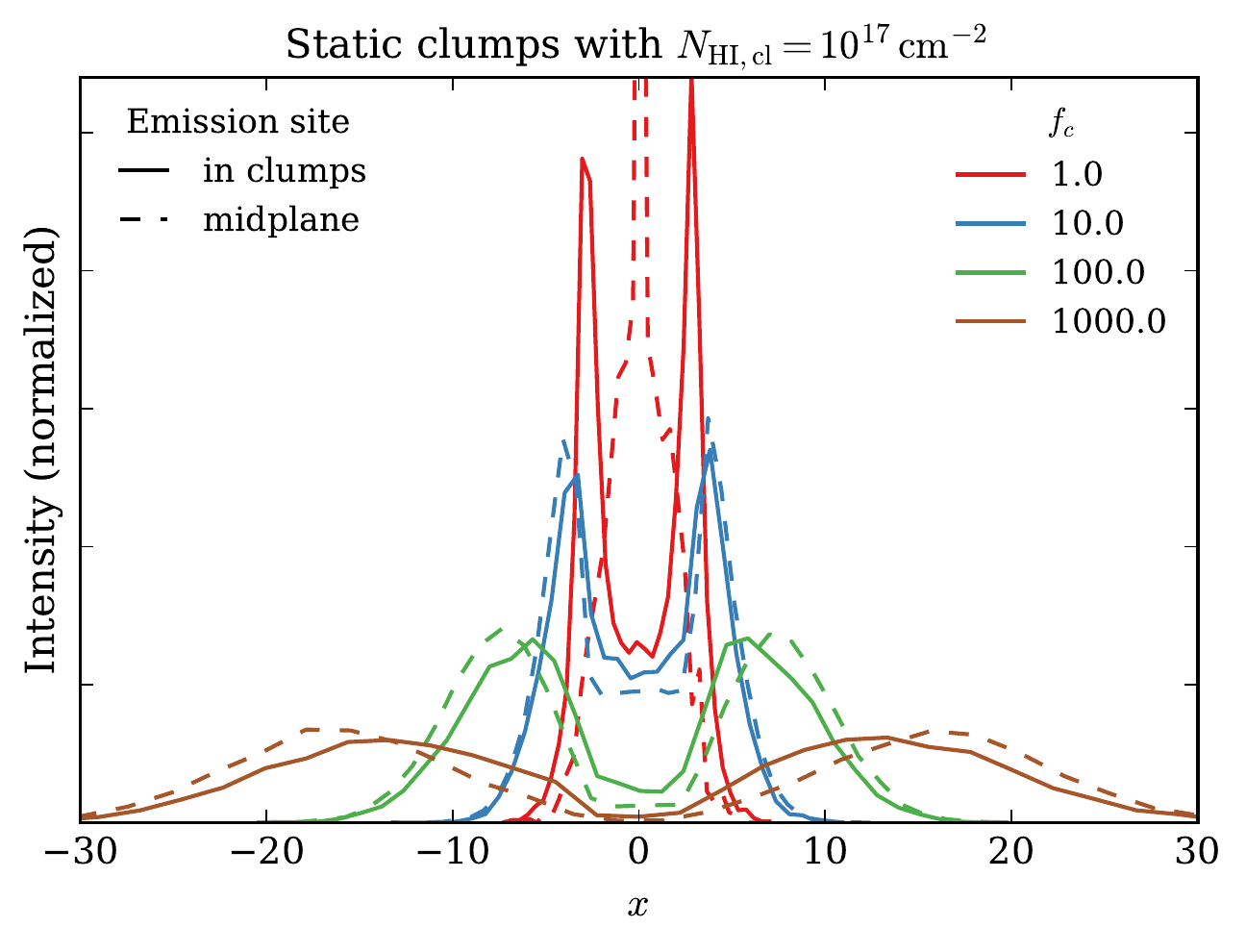}
  \caption{\Lya spectra for a constant clump column density $N_{\rm HI,cl}=10^{17}\cm^{-2}$ and three values of $\fc$. The \textit{solid lines} mark the spectral shape with emission inside the clumps whereas the \textit{dashed lines} show as comparison the spectra obtained from midplane emission.}
  \label{fig:spectra_clumpy-emission}
\end{figure}

\begin{figure}
  \centering
  \includegraphics[width=.95\linewidth]{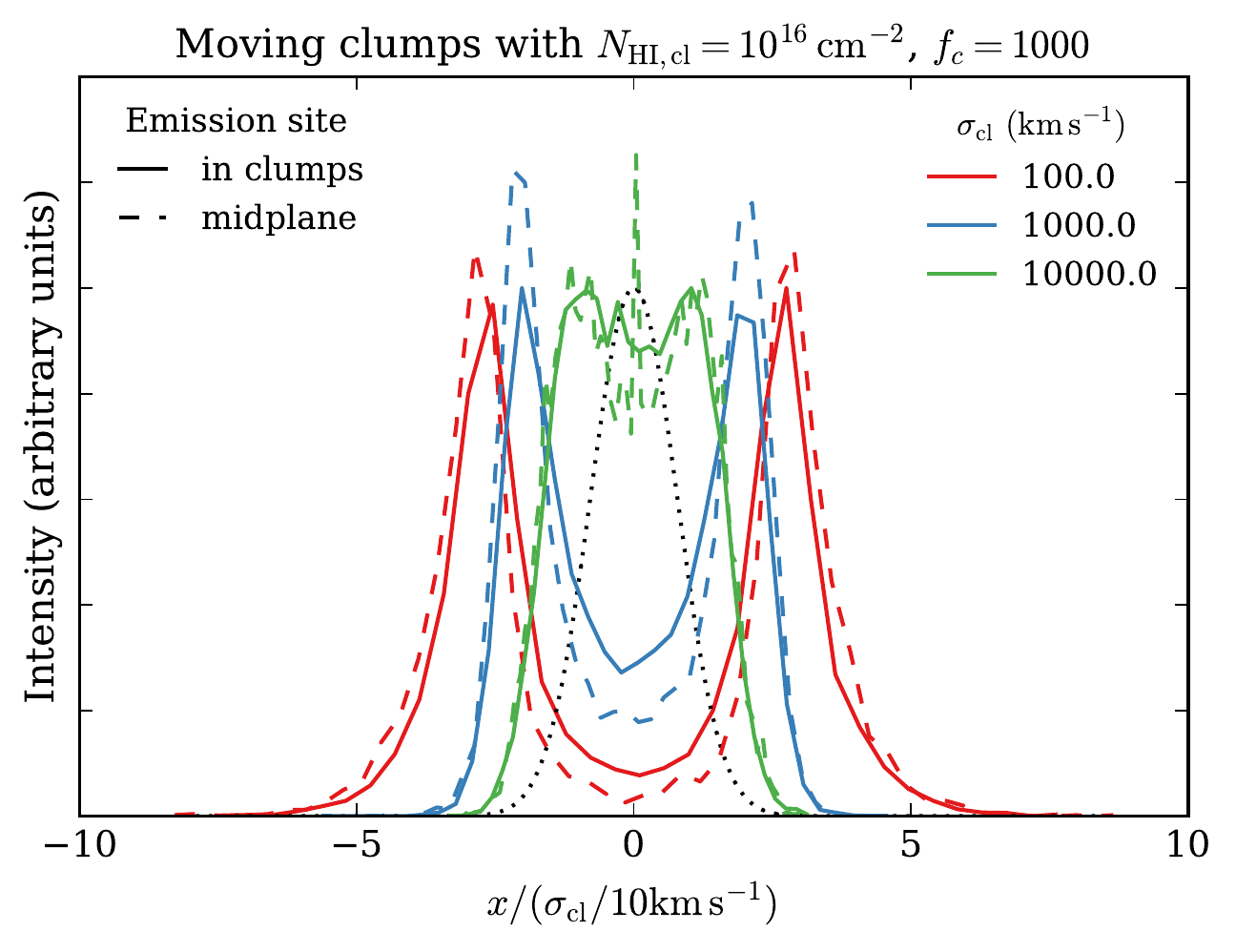}
  \caption{\Lya spectra for a constant geometry with $N_{\rm HI,cl}=10^{16}\cm^{-2}$, $\fc=1000$ and three values of $\sigma_{\cl}$. The \textit{solid lines} mark the spectral shape with emission inside the clumps whereas the \textit{dashed lines} show as comparison the spectra obtained from midplane emission. Please note that for presentation purposes we rescaled the $x$-axis according to the value of $\sigma_{\cl}$. The \textit{black dotted line} shows the intrinsic spectrum which has the same width for all $\sigma_{\cl}$ due to the rescaling.}
  \label{fig:spectra_clumpy-emission_moving}
\end{figure}

\subsection{Emission within the clumps}
\label{sec:clump-emission}
In this section we study the effect of emission originating from inside the clumps. This case resembles \Lya production due to cooling in the inner parts of the clumps, or to recombination events in the outer layer of (self-shielding) clumps caused by an external ionizing source. The latter is sometimes referred to as fluorescence. In both cases, \Lya are produced in the reference frame of the clumps, and experience an initial optical depth before entering the inter-clump medium -- both effects shape the `intrinsic' spectrum.

Fig.~\ref{fig:spectra_clumpy-emission} compares some spectra with starting position inside the clumps to the ones previously presented, i.e., with starting position at midplane. The clumps in this case possess a column density of $N_{\rm HI, cl}= 10^{17}\cm^{-2}$, which means they are optically thick to \Lya radiation. The effect of this can be seen best in the spectrum with $\fc = 1$ (red curve in Fig.~\ref{fig:spectra_clumpy-emission}) which contrasts a double peak profile due to the escape from the clump to the single peaked profile from the random walk process between the clumps. For greater values of $\fc$, however, this `initial feature' gets washed out from the scatterings off subsequent clumps and the spectra are independent of the emission site. 

In Fig.~\ref{fig:spectra_clumpy-emission_moving} we show a similar plot for moving clumps. Note that in this case the photons' frequencies are rescaled according to the value of $\sigma_{\cl}$ due to presentation purposes. This means that in the spectra for $\sigma_{\cl}=10^4\kms$ (shown in purple) are with a full width at half maximum (FWHM) of $\Delta x\sim 5000$ the widest of the presented spectra. 
As previously for the large $\fc\gg 1$ cases, the spectra with the emission sites within the clumps resemble closely  the ones with emission sites in the midplane. The only difference is that the latter are slightly wider and possess a smaller flux at line center which is simply due to the fact that a number of clumps are located at the boundary of the slab. This is encouraging as it shows that our results are quite general, that is, not dependent on the exact emission site. However, a small caveat is that for spherical geometries most clumps are located at large radii which might make this setup more sensitive to in-clump emission (on the other hand, the outermost clumps might emit less \Lya photons as some are `shadowed' by clumps closer to the ionizing source).

\subsection{Dusty clumps}
\label{sec:dust-within-clumps}
When placing absorbing dust in the clumps -- which we characterize by the all-absorbing dust optical depth $\tau_{\cl, d}$ -- \Lya can be destroyed leading to an escape fraction $f_{\rm esc}\le 1$. Interestingly, in clumpy medium the \Lya escape fraction might be larger than the continuum one as predicted by \citet{Neufeld1991}.
This `Neufeld effect' is due to the fact that \Lya photons may `surface scatter' off the neutral clumps, thus, effectively shielding the dust from them. Therefore, one expects the observed \Lya equivalent widths to be potentially much larger than the intrinsic ones. \citet{Hansen2005} characterized this effect more systematically using \Lya radiative transfer simulations for a wide range of parameters.
Building upon their work, \citet{Laursen2012} found, however, that in a part of the parameter space which they tried to constrain by observations the boosting vanishes. Specifically, out of their $4\times 10^3$ models only a few percent showed an equivalent width boost \citep[see also ][for a study of the `Neufeld effect' in clumpy shells]{Duval2013}. \citet{Laursen2012} thus concluded ``consider the Neufeld model to be an extremely unlikely reason for the observed high EWs''. All these studies focused on values of $\fc \sim 1$ and we want to re-visit \Lya escape in clumpy medium with several orders of magnitude greater covering factors. Thus, it is not entirely clear from the literature whether radiative transfer effects from clumpy media can explain the extreme equivalent width measurements observed in some galaxies. However, \citet{Laursen2012} identified some criteria which have to be fulfilled such as relatively slowly moving clumps with high dust optical depths.

Instead of re-running the radiative transfer simulations for various dust contents, we use the information of the hydrogen column density `seen' by each photon package to compute the \Lya escape fraction as in \citet{Gronke2015} which yields an escape fraction for each photon package that is
\begin{equation}
f_{{\rm esc,} i} = \exp\left[-\frac{\hat N_{{\rm HI}, i}}{N_{\rm HI, cl}} \tau_{\rm d, cl}\right]\;.
\end{equation}
Here, $\hat N_{{\rm HI}, i}$ is the column density experienced by photon package $i$. Given $f_{{\rm esc}, i}$ for a certain setup one can now obtain \textit{(a)} the overall \Lya escape fraction as the average of $f_{{\rm esc}, i}$, and \textit{(b)} the spectral shape altered through dust by simply assigning each photon package the weight $f_{{\rm esc}, i}$ when assembling the spectrum.

\begin{figure}
  \centering
  \includegraphics[width=.95\linewidth]{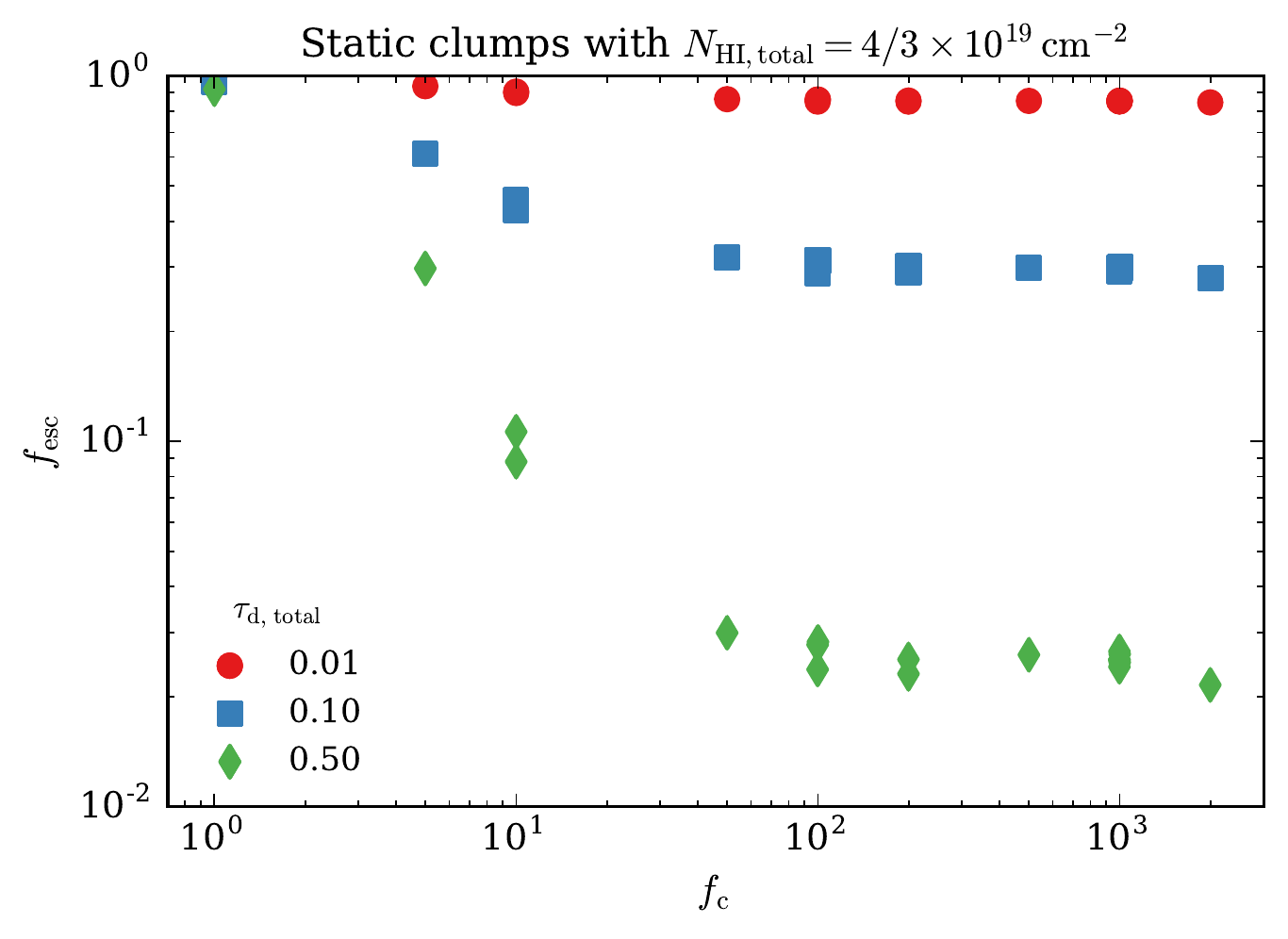}
  \caption{\Lya escape fraction versus $\fc$ for a fixed total hydrogen column density $N_{\rm HI, total}$ and dust optical depth $\tau_{\mathrm{d, total}}$.}
  \label{fig:fesc_static_simple}
\end{figure}

In Fig.~\ref{fig:fesc_static_simple} we plot the \Lya escape fraction versus $\fc$ for a constant total dust and hydrogen number content. For all three values of $\tau_{\mathrm{d, total}}$ displayed a similar trend is visible: with increasing \fc, first a approximately linear fall off in escape fraction before a flattening occurs, that is, $f_{\mrm{esc}}\sim \mrm{const.}$ for $\fc \gtrsim 40$.
Interestingly, the position of this threshold is independent of $\tau_{\mrm{d,total}}$ which hints towards a origin in the nature of the radiative transfer.
The flattening occurs at the boundary between the boundary between the `free streaming' and `homogeneous' regime because in the former the probability of absorption is proportional to the number of clump interactions (and, thus, \fc) whereas in the latter the escape fraction is set by the total dust content only and does not grow further with \fc. We discuss this phenomenon in more detail in \S~\ref{sec:escape-fractions-ew}.

\begin{figure}
  \centering
  \includegraphics[width=.95\linewidth]{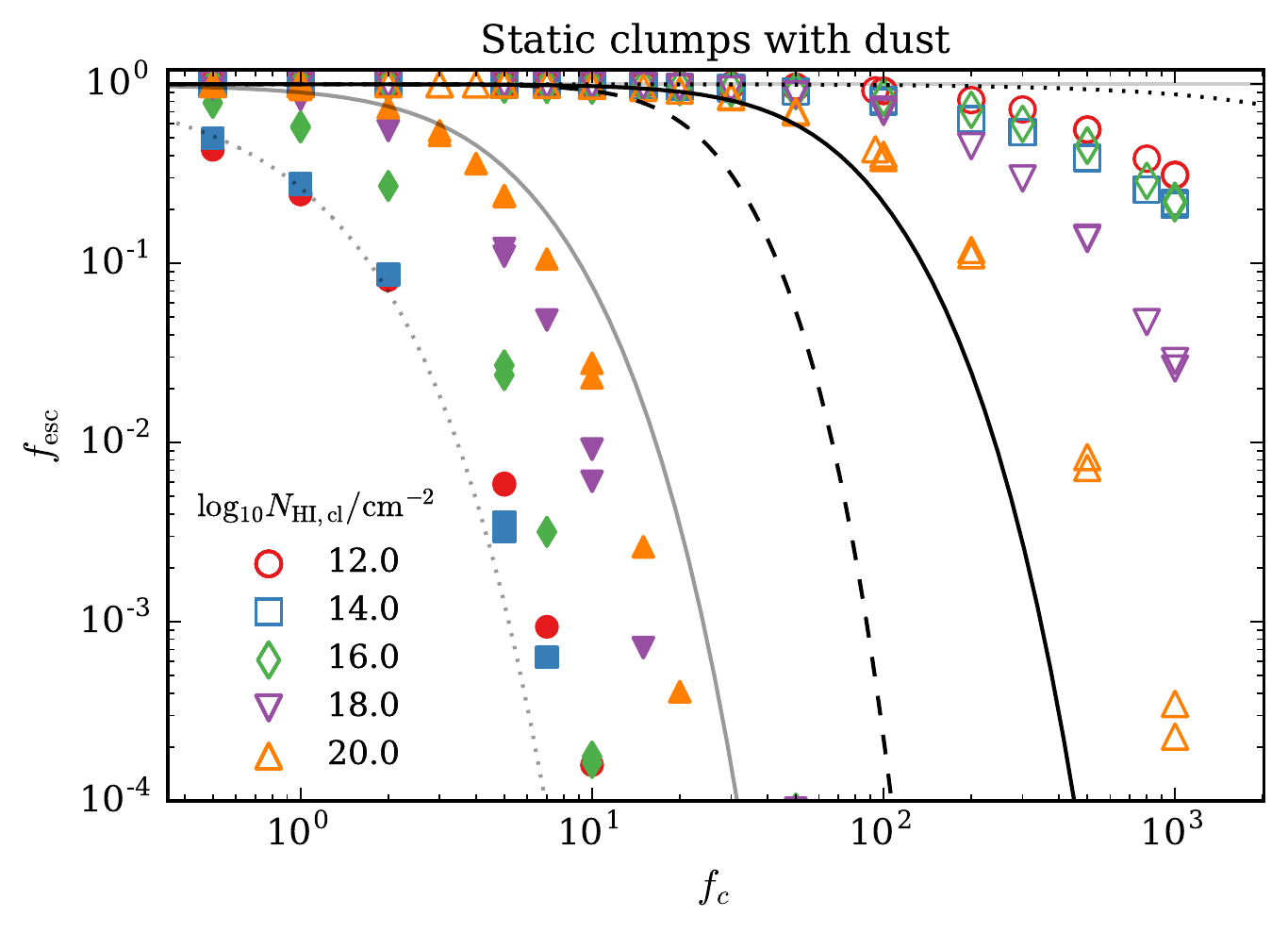}
  \caption{\Lya escape fraction versus $\fc$ for various values of $N_{\rm HI, cl}$, $\tau_{\rm d, cl}=1$ and $\tau_{\rm d, cl}=10^{-4}$ (filled and unfilled symbols, respectively). The black [grey] curves show some analytic curves for $\tau_{\rm d, cl} = 10^{-4}$ [$\tau_{\rm d, cl} = 1$]. The solid curve shows the \citet{Hansen2005} formula as given by Eq.~\eqref{eq:fescHO06}, the dashed line is the escape fraction from a homogeneous slab with $N_{\rm HI} = 4/3 \fc 10^{22}\cm^{-2}$ as given by \citet{Neufeld1990}, and the dotted line is simply $\exp(-4/3 \fc \tau_{\rm d, cl})$ symbolizing a continuum escape fraction.}
  \label{fig:fesc_static}
\end{figure}

An implication of the respective escape fractions of the two regimes is visible in Fig.~\ref{fig:fesc_static}. Here we show several values of $N_{\rm HI, cl}$ for the static setup using $\tau_{\rm d, cl}=10^{-4}$ (empty symbols) and $\tau_{\rm d, cl}=1$ (filled symbols) which correspond to metallicities of $Z/Z_\odot = 0.63\left(\tau_d / 10^{-4}\right)\left(10^{17}\cm^{-2} / N_{\HI,\cl}\right)$ \citep{Pei1992,Laursen2009} which reaches clearly unrealistic values. However, as in this paper we're interested in the fundamental impact of the individual parameters we also study these extreme value. 
Also shown in Fig.~\ref{fig:fesc_static} (with a black [grey] solid line for the low [high] dust content) is the proposed analytic solution for $f_{\rm esc}$ by \citet{Hansen2005}
\begin{equation}
f_{\rm esc}^{\rm HO06} = 1/{\rm cosh}(\sqrt{2 N_{\cl}\epsilon})\;,
\label{eq:fescHO06}
\end{equation}
where for $N_{\cl}$ we used Eq.~\eqref{eq:N_cl_generic} (with $(a_1,\,b_1) = (3/2,\,2)$ as found in \S~\ref{sec:static-case}) and for the clump albedo (i.e., the fraction of incoming photons which are reflected) $\epsilon$ we adopted a value of $c_1 (1 - e^{-\tau_{\rm d,cl}})$ with $c_1 = 1.6$ [$c_1=0.06$] to match the $N_{\rm HI, cl}=10^{22}\cm^{-2}$ data points for $\tau_{\rm d, cl}=10^{-4}$ [$\tau_{\rm d, cl}=1$].
Note, that the behaviour for the low and high dust contents is quite different. On the one hand, the escape fractions versus $N_{\rm HI,cl}$ scales for $\tau_{\rm d,cl}=1$ (filled symbols in Fig.~\ref{fig:fesc_static}) as predicted by \citet{Hansen2005} in their `surface scatter' approximation, that is, a larger clump hydrogen column density `shields' the dust better from the \Lya photons and thus prevents more efficiently their destruction. On the other hand, however, this is not the case for the low-dust scenario presented in Fig.~\ref{fig:fesc_static} (with unfilled symbols) where a larger value of $N_{\rm HI, cl}$ implies a lower $f_{\rm esc}$. This is due to the fact that here the dust optical depth through all the clumps (shown in the black dotted line in Fig.~\ref{fig:fesc_static}) is lower than the accumulated one through the subsequent random-walk clump encounters (black solid line), i.e, $\exp(-4/3 \fc \tau_{\rm d, cl}) \lesssim f_{\rm esc}^{\rm HO06}$.  Consequently, configurations in the `free-streaming' regime can possess enhanced \Lya escape fractions compared to the `random walk' regime (see \S~\ref{sec:escape-fractions-ew} for a more detailed discussion). 
Still, both cases possess (much) larger escape fractions than a homogeneous slab which is shown in Fig.~\ref{fig:fesc_static} with a black dashed line. Here, we use the derived escape fraction by \citet{Neufeld1990} with $N_{\rm HI}=4/3\times \fc 10^{22}\cm^{-2}$ and $\tau_{\rm d} = 4/3 \fc \tau_{\rm d, cl}$ -- i.e., with equal column densities as in the $N_{\rm HI, cl}=10^{22}\cm^{-2}$ case.

\begin{figure}
  \centering
  \includegraphics[width=.95\linewidth]{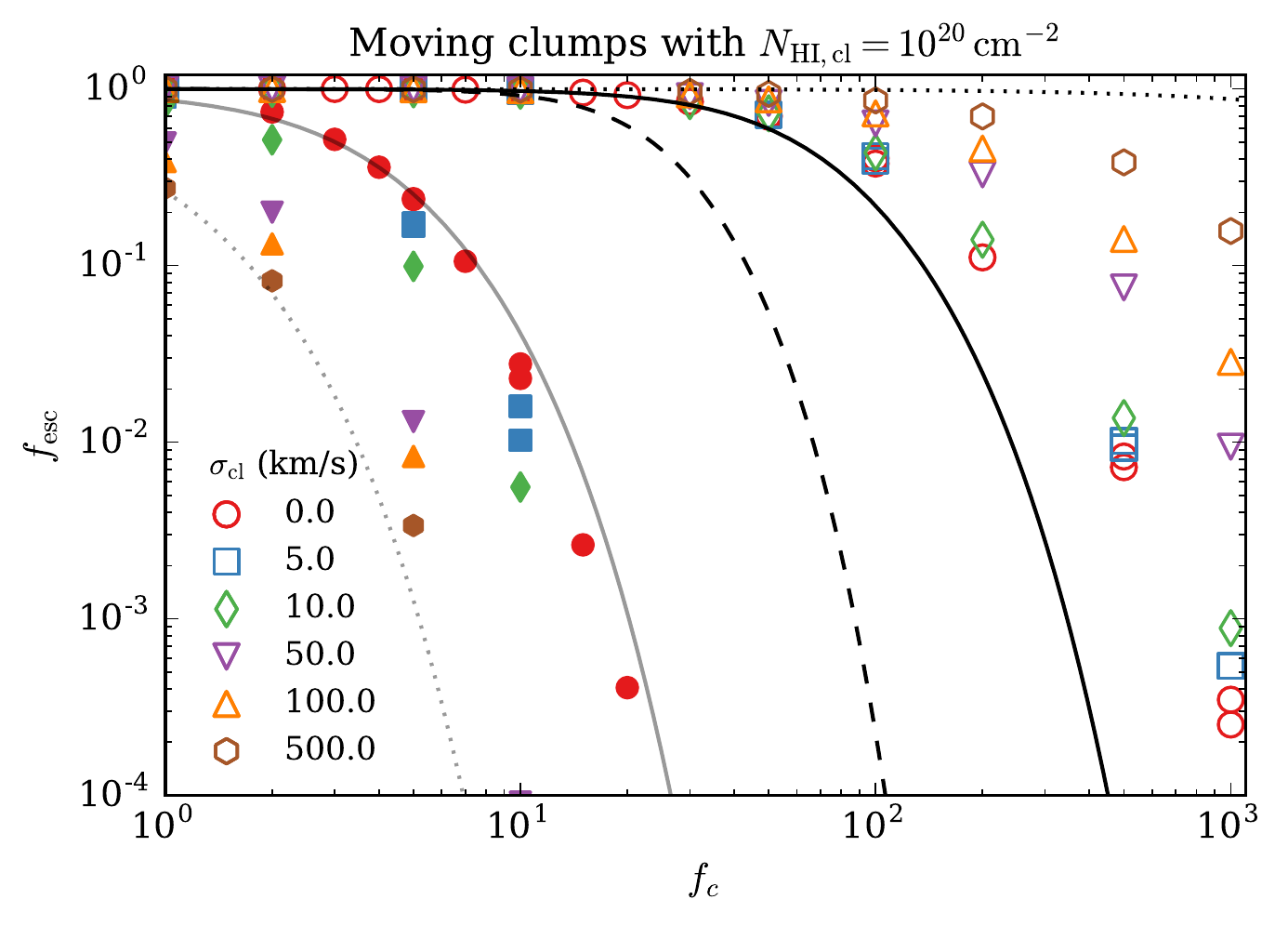}
  \caption{\Lya escape fraction versus $\fc$ for  fixed clumps with $\tau_{\rm d, cl}=10^{-4}$, $N_{\rm HI, cl}=10^{20}\cm^{-2}$ and various values of $\sigma_{\cl}$. The curves are the same as in Fig.~\ref{fig:fesc_static} for comparison.}
  \label{fig:fesc_moving}
\end{figure}

The same quantity, i.e., $f_{\rm esc}$ versus $\fc$, for the case of randomly moving clumps is plotted in Fig.~\ref{fig:fesc_moving}. As previously, the escape fraction departures from the curve given by Eq.~\eqref{eq:fescHO06} for $\fc \gtrsim \fccrit$. The lower number of clump encounters in this regime leads to a significantly higher escape fraction, e.g., $f_{\rm esc}\sim 10^{-1}$ for $\sigma_{\cl}=500\kms$.

\section{Discussion}
\label{sec:discussion}
In this section, we will discuss our results in the light of the various escape regimes discussed in Sec.~\ref{sec:analyt-cons-new} (\S~\ref{sec:regimes-clumpy-model}). Furthermore, we analyze what implications our results have for `\Lya equivalent width boosting' (in \S~\ref{sec:escape-fractions-ew}), and make the connection to observational results (of `shell-model' fitting; \S~\ref{sec:clumpy-solid-conn}) as well as to radiative transfer results through hydrodynamical simulations (\S~\ref{sec:impl-lya-radi}).

\subsection{The regimes of the clumpy model}
\label{sec:regimes-clumpy-model}
\begin{figure}
  \centering
  \includegraphics[width=.95\linewidth]{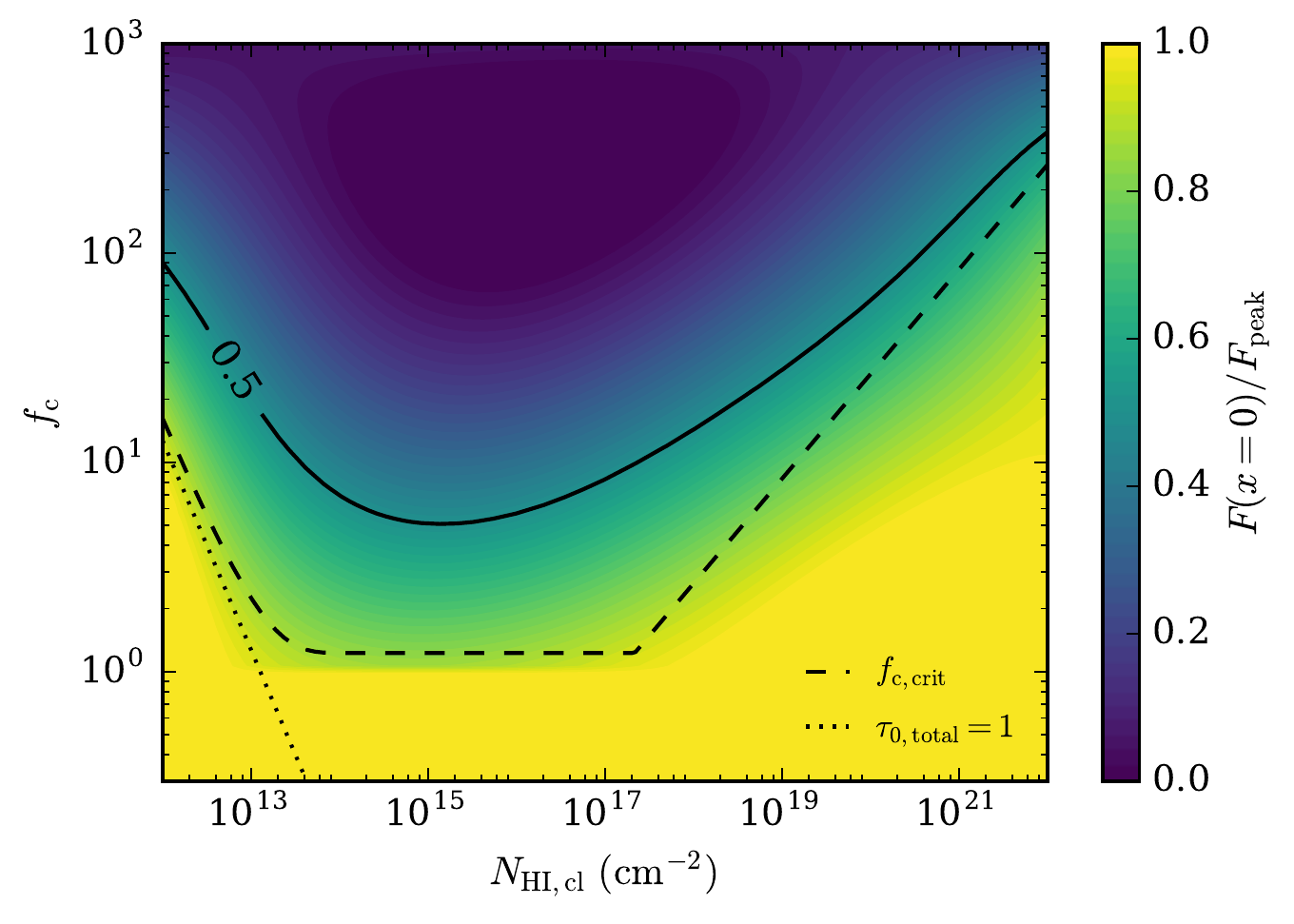}
  \caption{Overview of the different regimes for the static ($\sigma_{\mathrm{cl}}=v_{\mathrm{max}}=0$) setup. The color coding shows our (interpolated) numerical results in terms of the flux at line center divided by the peak flux of the spectrum, i.e., a value of $\sim 0$ [$\sim 1$] quantifies a double [single] peaked spectrum. Specifically this quantity is $1/2$ at the \textit{solid line}. The \textit{dashed} line marks the $\fccrit$ (Eq.~\eqref{eq:fccrit}), and the \textit{dotted line} is the boundary to the low-density regime ($\tau_{0, \mathrm{total}}=1$).}
  \label{fig:result_overview}
\end{figure}

Fig.~\ref{fig:result_overview} summarizes our findings for the static case. Here, color shows the flux at line center expressed in units of flux at the peak of the spectra $F(x=0)/F_{\rm peak}$. This measure is $\sim 1$ for a single peaked spectra and is less for double peaked spectra; a value of $\sim 0$ corresponds to an optically thick, `slab like' spectrum. We highlighted the dividing value of $F(x=0)/F_{\rm peak}=1/2$ specifically.

Also visible in Fig.~\ref{fig:result_overview} are the three regimes described Sec.~\ref{sec:analyt-cons-new}, along with our analytic estimates. They can be summarized as follows:
\begin{itemize}
\item \textit{Optically thin regime.} For an overall optical depth $\tau_{0, \mathrm{total}}=4/3 \fc\tau_{\rm 0, cl}\lesssim 1$ the $N_{\cl}-\fc$ scaling is shallow and the emergent spectra are single peaked. The dotted line in Fig.~\ref{fig:result_overview} mark this boundary.
\item \textit{Homogeneous regime.} If not in the `optically thin regime', for $\fc \gtrsim f_{\rm c,crit}$ we found also a shallower $N_{\cl}-\fc$ scaling than \citet{Hansen2005}. This is due to the preferential escape in an optically thick medium through single excursion -- which causes broad, double peaked spectra. Fig.~\ref{fig:result_overview} shows $f_{\rm c,crit}$ as a function of $N_{\rm HI,cl}$ as the dashed line.
Note that above this line we find $F(x=0)/F_{\rm peak}\rightarrow 0$ denoting double peaked spectra as predicted. Similarly, below this line the numerical results show single-peaked spectra.
\item \textit{Random-walk regime.} For optically thick clumps and $f_{\rm c}\lesssim \fccrit$ we recovered the results of \citet{Hansen2005}, i.e., $N_{\cl}\propto \fc^2$ and single peaked spectra due to a surface-scattering escape of the photons.
\end{itemize}

As noted in \S~\ref{sec:divis-betw-regim} these regimes break down for $\fc\lesssim 3$ -- an area of the parameter space which has previously by studied by \citet{Hansen2005,Laursen2012,Gronke2016a} -- where the probability of not finding a clump in a certain sightline is non-negligible \citep[this allows for non-zero ionizing photon escape fraction, see ][]{DijkstraLyaLyC2016}. 

\begin{figure*}
  \centering
  \includegraphics[width=.95\linewidth]{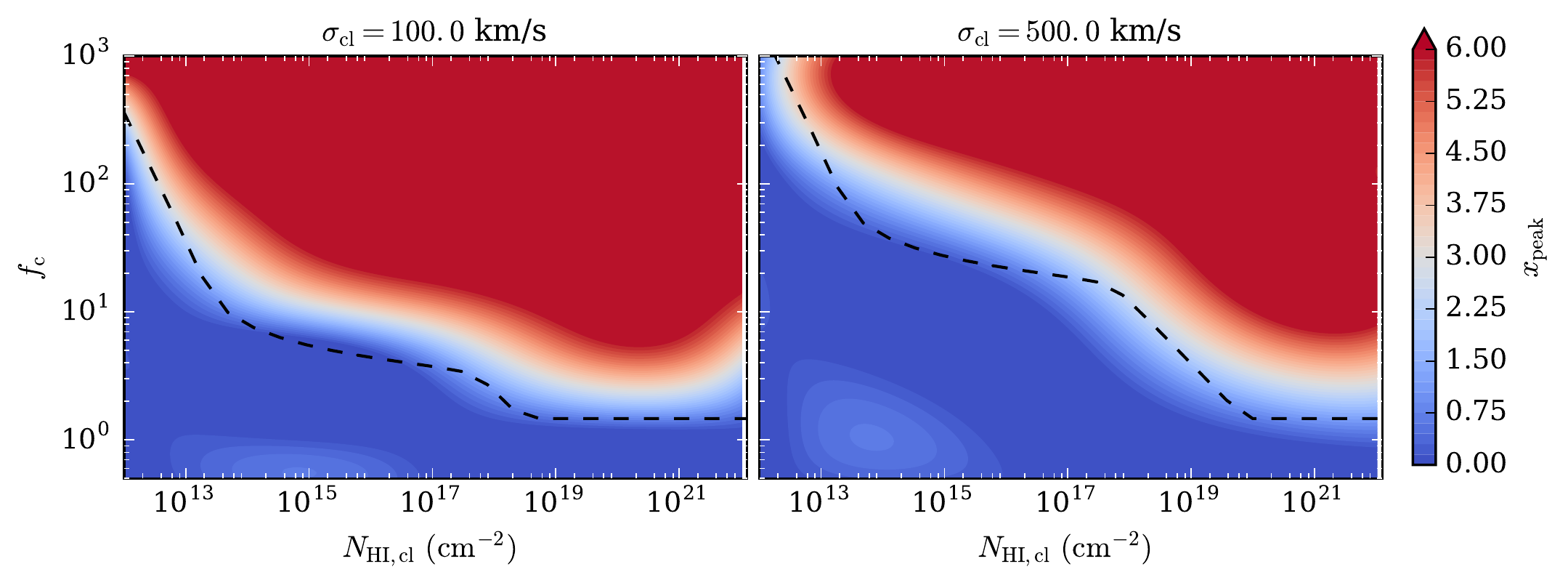}
  \caption{Overview of the $\fc$-$N_{\HI,\cl}$-plane with moving clumps for two different values of $\sigma_{\cl}$. The color coding shows the spectral peak position $x_{\mathrm{peak}}$ (truncated at $x_{\mathrm{peak}}=6$). 
 The \textit{dashed lines} show \fccrit in the moving case (\S~\ref{sec:non-static-case}), i.e., below this line the velocity distribution of clumps is not sampled well.
}
  \label{fig:result_overview_moving}
\end{figure*}

Fig.~\ref{fig:result_overview_moving} displays the transition from double- to single-peaked spectra for randomly moving clumps. The color coding shows in this case the peak position of the spectrum with white being $x_{\mathrm{peak}}\sim 3$, that is, when the peak position moves outside the core of the line\footnote{We used an alternative criterion because for larger $\sigma_{\cl}$ the spectra can be very broad, and thus, $F(x=0)/F_{\rm peak}$ becomes noise dominated. However, both measures can be used to distinguish between single- and double-peaked.}. 
For faster clumps, this boundary moves to greater values of \fc, making it more likely to obtain a single-peaked spectrum (at line center).
The black dashed lines in Fig.~\ref{fig:result_overview_moving} denotes \fccrit from Eq.~\eqref{eq:fccrit_moving}\footnote{In fact, we used the exact functional form for $\sigma_{\HI}(x)$ and did not resort to the approximation as in Eq.~\eqref{eq:fccrit_moving} which yields a slightly better fit to the data..} -- in other words below this line the velocity space is not well sampled and allows photons at line center to escape.

\begin{figure}
  \centering
  \includegraphics[width=.95\linewidth]{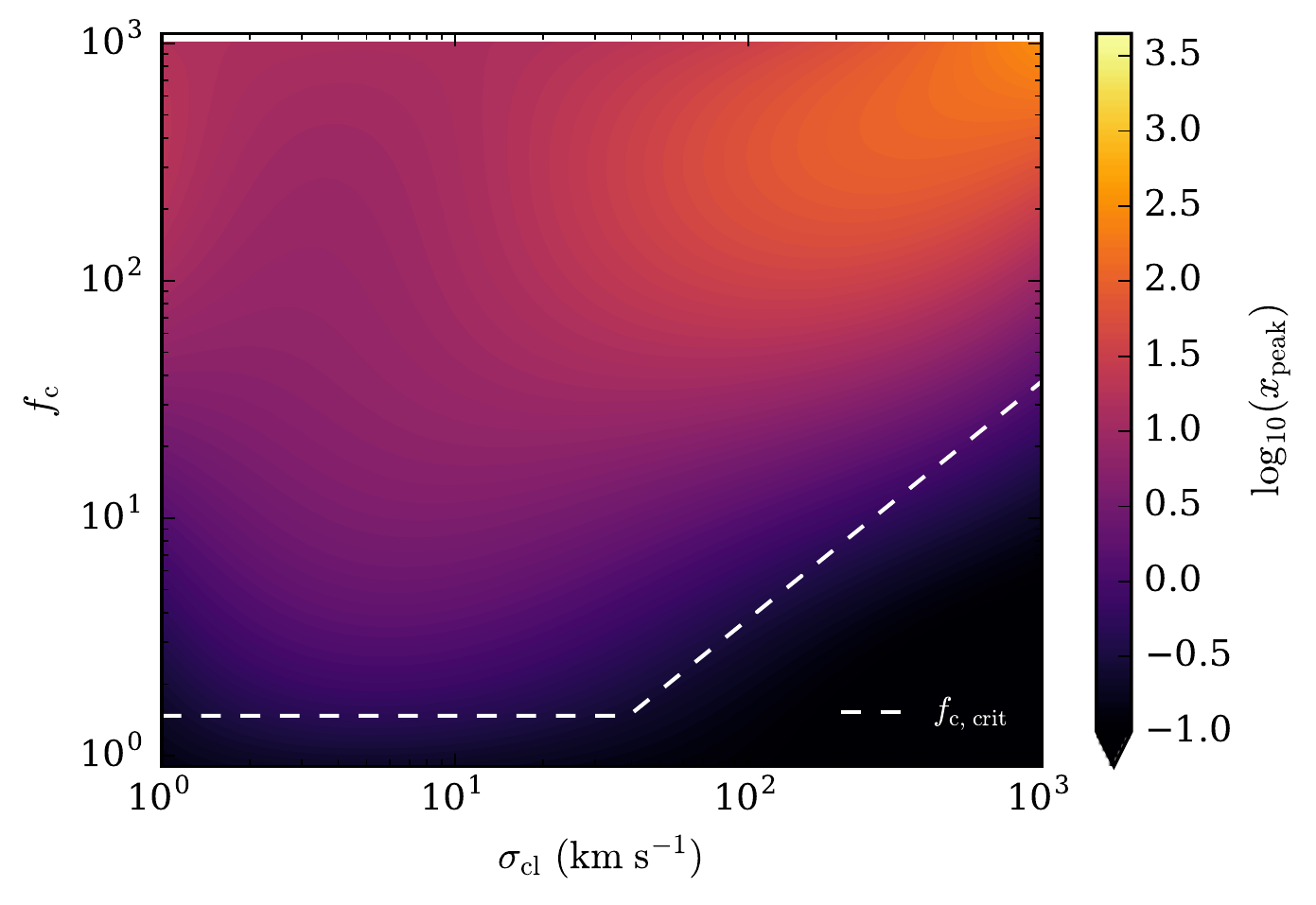}
  \caption{Overview of the spectral shape for $N_{\HI,\cl} = 10^{17}\cm^{-2}$, the clump column density predicted by \citet{McCourt2016}. The color coding denotes (the log of) the peak position $x_{\mathrm{peak}}$, i.e., low values (in black) represent a single peaked feature. The white dashed line is again the \fccrit boundary in the moving case.}
  \label{fig:overview_NHI17}
\end{figure}

The same line is marked also in Fig.~\ref{fig:overview_NHI17} where we focus on the clump column density $N_{\HI, \cl}=10^{17}\cm^{-2}$ as predicted by `shattering' \citep{McCourt2016}. Here the peak position (in log scale) is color coded as a function of covering factor and clump velocity dispersion. 
For large values of $\sigma_{\cl}$, the transition to double peaked spectra occurs at a larger covering fraction, since more clumps are required to sample the broader velocity distribution.
Below this threshold, we see a single-peaked spectrum from photons which escape through holes in velocity space.

Fig.~\ref{fig:overview_outflow_asymmetry} shows this increase in asymmetry with greater \fc (for fixed outflow velocity and total column density of $N_{\HI, \mathrm{total}}=4/3\times 10^{19}\cm^{-2}$). Here, the color corresponds to the asymmetry of the spectra which we define as in \S~\ref{sec:outflows} to be the ratio of the integrated blue over the red flux minus one. In Fig.~\ref{fig:overview_outflow_asymmetry} we also mark graphically the conditions for homogeneous escape discussed in \S~\ref{sec:non-static-case}, that is, that the adjacent clump is optically thick ($4/3 \tau_\cl(x_{\mathrm{next}}) \gtrsim 1$ with $x_{\mathrm{next}} \equiv v_{\mathrm{max}} / (\fc v_{\mathrm{th}})$), and that the initial scatterings occur in the core of the line ($x_{\mathrm{next}} < x_*$). If both conditions are fulfilled (and sufficient outflows are present, i.e., $v_{\mathrm{max}}\gtrsim 50\kms$), the emergent spectrum is asymmetric towards the red side (as visible from the red region in Fig.~\ref{fig:overview_outflow_asymmetry}).

\subsection{Escape fractions and EW boosting}
\label{sec:escape-fractions-ew}

The different regimes for \Lya radiative transfer through a multi-phase gas have different implications for the \Lya escape fraction when dust in present within the clumps. 
\begin{itemize}
\item Inside the `optically thin regime' ($\tau_{0, \mrm{total}}\lesssim 1$), the \Lya escape fraction is equal to the continuum one as \Lya photons stream through all the clumps and are affected by the dust content within them. Hence, $f_{\mrm{esc}}\approx \exp(-\tau_{\mrm{d, total}})$. This can be seen for the $N_{\HI,\cl}\lesssim 10^{14}\cm^{-2}$, $\fc \lesssim\fccrit$ data points in Fig.~\ref{fig:fesc_static}.
\item In the `random-walk regime', we confirm the escape fraction given by \citet{Hansen2005} (see Eq.~\eqref{eq:fescHO06}) (apart from geometrical pre-factors). In this regime the governing quantity for the escape fraction is $\epsilon$, i.e., the absorption probability per clump interaction and the number of clumps encountered $N_\cl$. In this regime, the latter is merely a function of \fc (see Eq.~\eqref{eq:N_cl_generic}) but $\epsilon$ depends non-trivially on $N_{\HI, \cl}$ and the clump movement due to variations in how deep the photons penetrate into the clumps. This is why there is some scatter in $f_{\mrm{esc}}$ in this regime; visible, for instance, in Fig.~\ref{fig:fesc_static}. This is the only regime where \Lya photons are shielded from dust, thus, allowing for `EW-boosting' \citep{Hansen2005}. That is, the ratio between the \Lya and UV escape fraction might be greater than unity.

\item Finally, in the `homogeneous regime', the behaviour is a combination of the above two behaviours. Initially, the photons will (on average) interact with $\sim N_\cl(\fccrit)$ clumps before diffusing to the line wings and escaping through free-streaming which leads to another $\sim \fc$ clump encounters (cf. Fig.~\ref{fig:nclumps_static}). Consequently, in this regime the escape fraction is approximately given by $f_{\mrm{esc}}\sim f_{\mrm{esc}}^{\mrm{HO06}}(\fccrit)e^{-\tau_{\mrm{d,total}}}$. This causes the flattening of $f_{\mrm{esc}}$ versus \fc in Fig.~\ref{fig:fesc_static_simple} as in this case $\tau_{\mrm{d, total}}$ is kept constant.
\end{itemize}

From the above considerations, one can see that the escape fraction
depends on several parameters and is therefore non-trivial to predict.
As a consequence, in \S~\ref{sec:dust-within-clumps}  we demonstrated
that the \Lya escape fraction may either increase \textit{or} decrease
with increasing metallicity which is $Z \propto \tau_d / n_{\HI}$ \citep{Pei1992,Laursen2009}, depending on the dust optical depth through an individual clump $\tau_{\mrm{d,cl}}$ (see the trends in the filled and unfilled symbols in
Figs.~\ref{fig:fesc_static} \& \ref{fig:fesc_moving}).
The controlling parameter is essentially the
ratio of absorption probability per surface scatter to the absorption
probability per clump passing.
Moreover, we have shown that \Lya escape fractions can be large, even
for large values of \fc.  Thus, we find homogeneous, `slab-like'
spectra can be observable even in models with significant dust content
(as is realistic; see \S~\ref{sec:clumpy-solid-conn} \&
\S~\ref{sec:impl-lya-radi}).

Regarding the EW boosting we found, one necessary requirement for the `Neufeld effect' to be active is that \Lya photons escape via surface scatterings off the clumps, i.e., in the `random walk' regime. This implies that the emergent spectrum is narrow, and single peaked at line center \citep[as already noted by ][]{Laursen2012} -- a clear observational signature for EW boosting to be active\footnote{Note however that if the line is narrow and concentrated on line center, that then the IGM can suppress the flux, as this is where we expect the IGM opacity to peak \citep[see, e.g.,][]{Laursen2011}.}. 

\subsection{Connection to a homogeneous medium}
\label{sec:clumpy-solid-conn}
Observed \Lya spectra can often successfully modeled using a simple model called the `shell-model' \citep[see, for instance,][]{Hashimoto2015,Karman2016_dl}. This shell-model consists of a central \Lya (and continuum) emitting region which is surrounded by a moving shell of hydrogen and dust \citep{Ahn2003MNRAS.340..863A,Verhamme2006A&A...460..397V}.  It is somewhat surprising that this simple, six-parameter model can account for the likely radiative transfer effects happening in the complex, multiphase medium of a variety of galaxies and their environments.
Since the shell-model is clearly very idealized, it is unclear what the extracted shell-model parameters mean physically. In \citet{Gronke2016a} we found that a simple one-to-one mapping between the shell-model parameters and the ones from a clumpy medium is not possible -- for the most part, the shell-model cannot reproduce the spectra emergent from a multiphase medium. This failure mostly results from the high fluxes at line-center from the multiphase simulations which are hard to obtain through radiative transfer through a uniform gas distribution (such as a shell).

\begin{figure*}
  \centering
  \includegraphics[width=.95\textwidth]{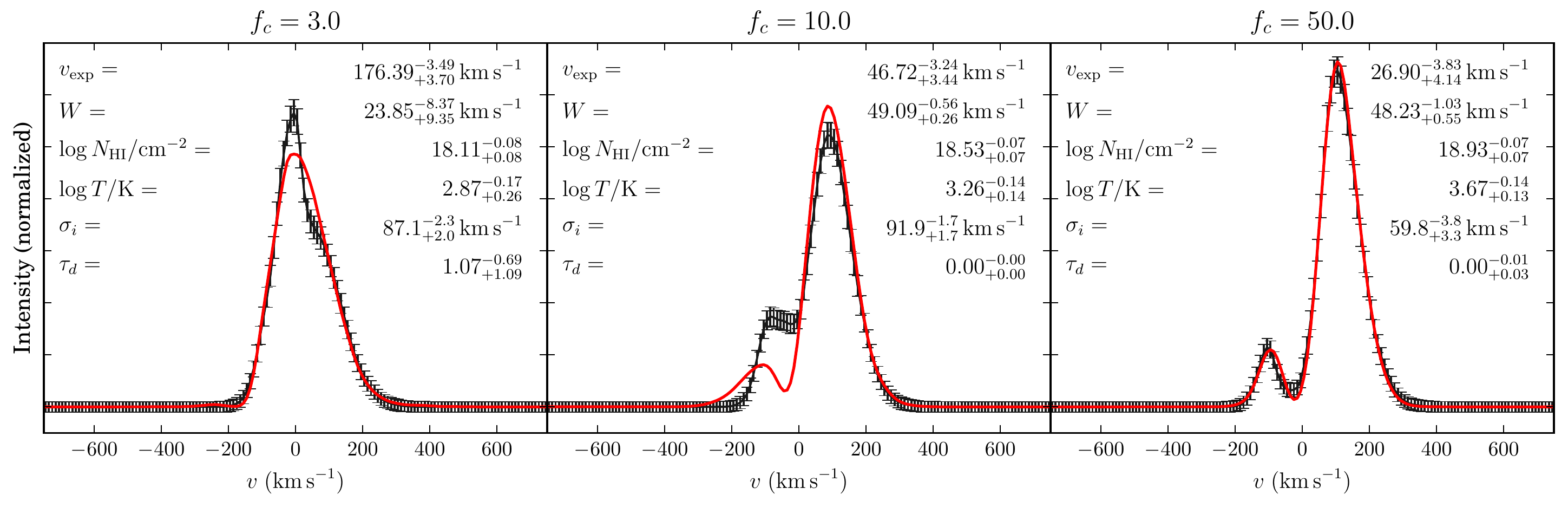}
  \caption{Fitting shell models to spectra of $\log N_{\HI, \mathrm{total}} / cm^{-2} \approx 19.1$ and $v_{\mathrm{max}}=50\kms$ and covering factors of $\fc = \{3,\,10,\,50\}$ (left to right panel). The black points and red lines show the spectra of the multiphase media, and the best fit shell-model spectra. See \S\ref{sec:clumpy-solid-conn} for details.}
  \label{fig:shell_model_fits}
\end{figure*}

However, in \citet{Gronke2016a} we restricted our analysis to covering factors of $\fc \sim \mathcal{O}(1)$ (and $\sigma_{\cl}\lesssim 100\kms$) -- i.e., the `random-walk' and `optically thin' regime. As we showed here, for (much) greater number of clouds the system approaches a `slab like' state which leads to, e.g., much lower fluxes at line center for the resulting \Lya spectrum. Hence, these multiphase spectra might be closer to observed ones.
Whether or not the shell-model parameters correspond to physical parameters of a clumpy medium with large $\fc$ is part of future work. However, our results show that for $\fc\gtrsim \fccrit$ the spectra are similar to a slab with the same column density. Furthermore, we fitted shell-models to three spectra originating from a clumpy medium with $N_{\HI, \mathrm{total}}=4/3\times 10^{19}\cm^{-2}$, $v_{\mathrm{max}}=50\kms$, and various covering factors. Prior to fitting, we smoothed the spectra using a Gaussian kernel with FWHM $W\sim 24\kms$.
Fig.~\ref{fig:shell_model_fits} shows the three spectra as well as the best-fit shell model spectra. The resulting shell-model parameters are also displayed in the figure. While the fits for $\fc = 3$ and $\fc = 10$ are rather poor and the recovered shell-model column densities are more than an order of magnitude off, the spectra for $\fc=50$ can be remarkably well recovered. Interestingly, here the shell column density is very close to the input value, and the recovered shell of $v_{\mathrm{exp}}\approx 25\kms$  outflow velocity corresponds to the mass weighted mean of the used Hubble-like outflow. Also, the dust content and to some extent the temperature of the gas are recovered. On the other hand, as the photons are injected at line center, the recovered widths of the intrinsic spectra ($\sigma_i$) are too large.
This may be to compensate the narrow coverage of the shell in velocity space. Note that a similar discrepancy of the intrinsic profile width is also found in the literature \citep[e.g.,][by comparing $\sigma_{\mathrm{i}}$ with the width of the H$\alpha$ line]{Yang2015} where it might also originate from radiative transfer effects.

All these points suggest that at least some of the shell-model parameters might have a true physical meaning. In this work, we provide an equally simple but physically meaningful model which serves as theoretical justification for the shell-model.
The full mapping from shell-model to parameters of a multiphase medium with large \fc is part of future work. However, from our single example it is already apparent that if an observed \Lya spectra can be modeled using a simple, homogeneous shell, one can -- and we encourage the reader to do so -- think instead about a fog of droplets (with $\fc \gg \fccrit$), which is more realistic given our knowledge about gas properties.

\begin{figure}
  \centering
  \includegraphics[width=0.95\linewidth]{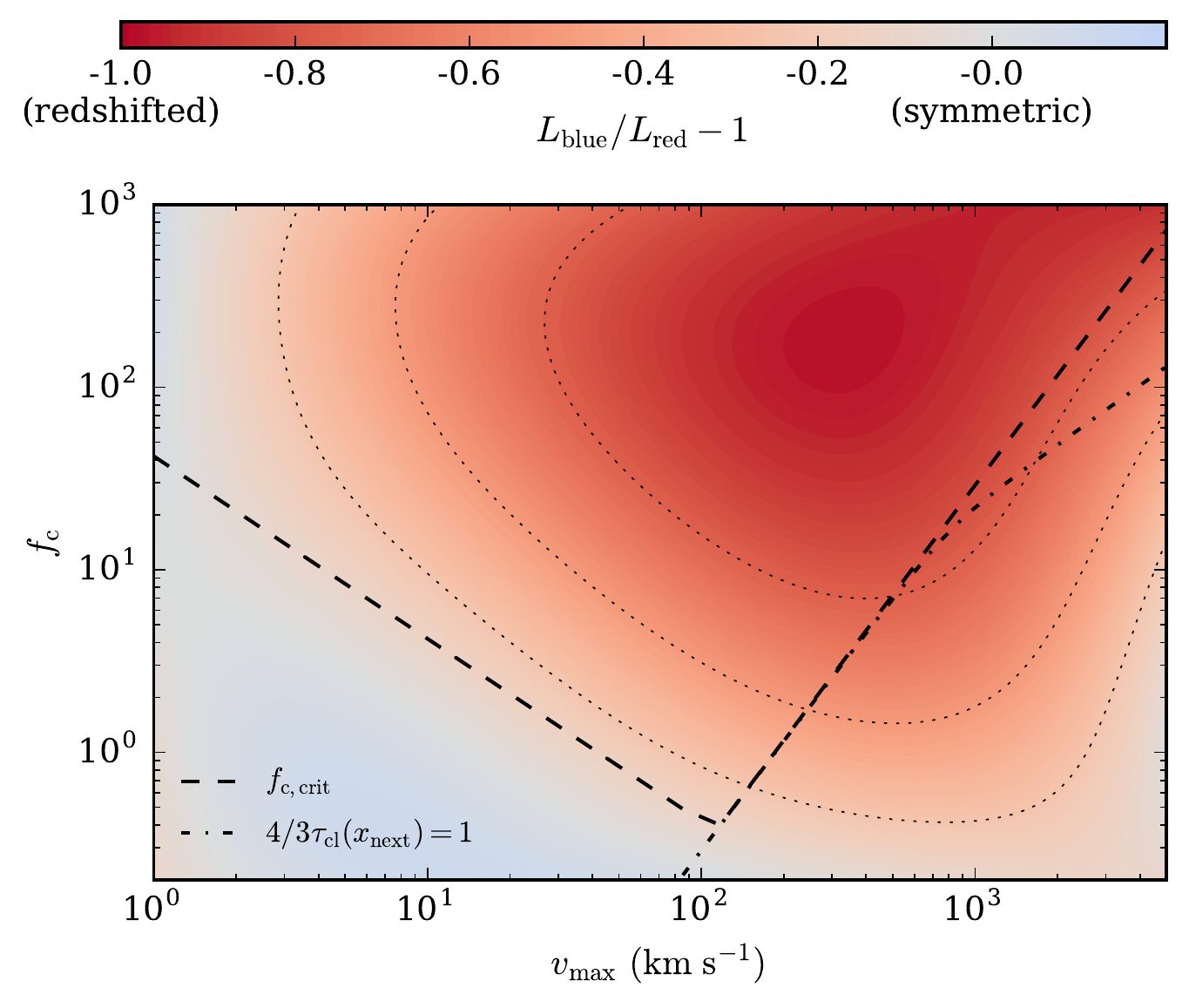}
  \caption{Asymmetry of the spectra (color coded) as a function of outflow velocity $v_{\mathrm{max}}$ and covering factor \fc for a fixed total column density of $N_{\HI, \mathrm{total}}=4/3\times 10^{19}\cm^{-2}$. As tracer of the asymmetry we chose to display the ratio of the integrated flux on the blue side ($x \ge 0$) of the line $L_{\mathrm{blue}}$ over the integrated red flux $L_{\mathrm{red}}$ minus one. This implies a value of $0$ (in white) corresponds to a symmetric spectrum whereas $-1$ (in dark red) to a spectrum where all flux is redward of line center. The contour lines highlight values of $(-0.75,\,-0.5,\,-0.25)$. Also shown are the \fccrit boundary (Eq.~\eqref{eq:fccrit_outflow}), and the more precise $4/3\tau_{\cl}(x_{\rm next})=1$ deviation. The relatively low values of \fccrit imply that large covering factors as predicted by \citet{McCourt2016} will lead to asymmetric spectra.}
  \label{fig:overview_outflow_asymmetry}
\end{figure}

\subsection{Implications for \textit{ab initio} \Lya radiative transfer simulations}
\label{sec:impl-lya-radi}
Our findings suggest a possible cause for the mismatch between observed \Lya spectra and the ones computed using snapshots of hydrodynamical simulations as input -- which are sometimes referred to \textit{ab initio} \Lya radiative transfer simulations.

Observed \Lya spectra from $z\sim 0$ to higher redshifts show several common features:
\begin{itemize}[itemsep=10pt]
\item A significant shift redwards of the main emitting peak. For instance, at $z\sim 2-3$ galaxies selected due to their strong \Lya emission as well as dropout-selected galaxies (LAEs and LBGs, respectively) show shifts of several hundred \kms\ \citep[e.g.,][]{Steidel2010a,Kulas2011,Erb2014,2014ApJ...791....3S,Trainor2015,Hashimoto2015}.
\item Asymmetric profiles with mostly stronger red than blue component. For instance, \citet{Erb2014} measured the median EW ratio $W_{\rm blue} / W_{\rm red}$ in their sample of $36$ LAEs at $z\sim 2-3$ to be $\sim 0.4$. This is consistent with the findings at $z\sim 0.2$ that also show a dominant red side (by a factor of a few)  \citep{Henry2015,Yang2015,Yang2017}.
\item There are roughly as many single as double peaked spectra. For instance, in the \Lya selected galaxy sample presented by \citet{Trainor2015} of $318$ LAEs at $z\sim 2.5-3$, $41\%$ show a double peaked spectra. This fraction agrees well with the  double peaked fraction of $45\%$ they found in the KBSS-MOSFIRE LBG sample \citep{2014ApJ...795..165S}, and the ones of other studies \cite[e.g.,][found ratios of $\sim 1/3$ and $1/2$, respectively]{Kulas2011,2012ApJ...751...29Y}. Note that only a small part of the double peaked profiles show a dominant blue peak which agrees with flux ratio discussed above. \citet{Trainor2015} quantifies these to be $\sim 10\%$ of the double peaked spectra.
\item For double peaked spectra, the flux in the `valley' between the peaks is small. Because of smoothing and resolution effects due to the observational aperture, measuring this quantity is challenging -- in particular for higher redshifts. However, at lower redshifts \citet{Yang2015} find in their sample of $12$ galaxies at $z\sim 0.2$ the flux ratio between the valley and the red peak to be $0.03^{+0.08}_{-0.02}$ and never greater than $0.27$.
Also the $14$ galaxies of the `Lyman-$\alpha$ reference sample' \citep[LARS; ][]{Ostlin2014} at $0.02 < z < 0.2$ have a flux ratio between the maximum and the minimum of $<0.1$, mostly even consistent with zero \citep{Rivera-thorsen2014}.
\end{itemize}

These findings seem to be in stark contrast to \Lya radiative transfer simulations that use a snapshot of a (high-resolution) hydro-dynamical simulation of a galaxy as input geometry
\citep[e.g.,][]{2006ApJ...645..792T,Laursen2007ApJ...657L..69L,Zheng2009,2011MNRAS.416.1723B,Verhamme2012,Behrens2014a,Smith2014,Trebitsch2016}. Due to computational cost, and probable directional dependence of the emergent spectrum \citep{Verhamme2012,Behrens2014a}, no statistical compilation of simulation-based spectra has yet been assembled. 
However, existing predicted spectra are generally too symmetric and / or possess a too high flux at line center. This is commonly attributed to 
\textit{(i)} CGM in combination with instrumental effects \citep[as discussed in ][]{Gronke2016a}, \textit{(ii)} radiative transfer effects  in close proximity to the origin of the photon, and / or
\textit{(iii)} IGM  absorption \citep{Dijkstra2007,Laursen2011}. 
All these arguments move the problem to different spatial location (in case of \textit{(ii)} even to a subgrid scale). However, the last solution cannot be universally invoked, especially at lower redshifts. For instance, \citet{Laursen2011} found that for $z\lesssim 3.5$ only $\lesssim 30\%$ of the sightlines show a full absorption feature which would lead to a low flux in the `valley'. 
Furthermore, while the IGM opacity increases with redshift, we see that the \Lya escape fraction from star-forming galaxies also increases with redshift \citep{Hayes2011,2011ApJ...736...31B,2013MNRAS.435.3333D}.
Both arguments strongly suggest that IGM absorption cannot be the dominant mechanism regulating the visibility of \Lya emission.

We have shown here that this discrepancy between observations and simulations can be understood easily. 
Simulations with Lagrangian-type techniques such as adaptive-mesh-refinement or smooth-particle-hydrodynamics (AMR and SPH, respectively) reach
their highest resolution in the densest regions such as the mid-plane
of the galaxy disk.  While future simulations will likely reach peak
resolutions approaching the $\sim0.1$\,pc scale we expect, we note that this \textit{still} won't capture clump formation
and evolution at large distances in the CGM, where the density and
resolution remains low. 
This means the clumps are unresolved and, thus, the covering factor per resolution element is less than unity -- compared to potentially hundreds as suggested theoretically by \citet{McCourt2016}. This lower $\fc$ (while keeping the column density and global structure unchanged) leads to a higher flux at line center (as shown in Fig.~\ref{fig:spectra_fc}), less asymmetric spectra (Fig.~\ref{fig:overview_outflow_asymmetry}), and in general more `unrealistic spectra' \citep[cf.][]{Gronke2016a}.
Therefore, small-scale structure in the CGM is crucial for modeling radiative transfer through the galaxy.
We expect that
direct simulation of the multiphase CGM will be essentially
impossible, precisely because it requires very high spatial
resolution, even in parts of the galaxy which are typically empty:
for example, a spatial resolution of $\sim0.1$\,pc in the outskirts of
a galaxy corresponds to a mass resolution of
$\sim{10}^{-10}$\,--\,$10^{-9}$ solar masses.
Instead, we propose to study \Lya radiative on the smallest scales, and then to use this knowledge as a sub-grid recipe.


\section{Conclusion}
\label{sec:conclusion}
Motivated by several observations and recent theoretical work by \citet{McCourt2016} we studied \Lya radiative transfer in an extremely clumpy medium, i.e., with large number of clumps per sightline (up to $\fc \sim 1000$). Our main findings on \Lya radiative transfer through clumpy media are:
\begin{itemize}[itemsep=10pt]
\item The behaviour of a multiphase medium depends strongly on the `clumpiness' of the system -- even when keeping the other parameters such as the total column density constant.
\item In particular, we identify a threshold above which \Lya photons escape preferentially via frequency excursion, i.e above which multiphase media affect \Lya as if they were homogeneous.
 This transition depends on clump column density and can be estimated analytically.
We found the threshold for the static case to be:
\begin{equation}
\fccrit \approx  \begin{dcases}
\frac{2 \sqrt{a_v \tau_{0, \cl}}}{3 \pi^{1/4}}  & \text{ for } \sqrt{3} a_v \tau_0 \gtrsim 19\\
\frac{1.65}{1-\e^{-\tau_{0,\cl}}}  & \text{ otherwise.}
\end{dcases}
\end{equation} 
\item The value of this threshold between clumpy- and homogeneous nature further depends on the clump kinematics in a way that can also be estimated analytically.
  If the clump motion is uncorrelated and Maxwellian, we find that the threshold is given by:
\begin{equation}
  \fccrit \approx \mathrm{max}\left( \frac{3\sigma_\cl}{2 v_{\rm th} \sqrt{\ln(4/3 \tau_{0,\cl})}},\; 1.5 \right).
\end{equation}
This is valid for sufficiently large clump motion, i.e., $\sigma_\cl \gtrsim v_{\mathrm{th}}$. For smaller values, the system approaches the static case above. Furthermore, we expect for large-scale correlations in velocity the transition to happen between these extreme cases. We will investigate this in a future study.
\item A similar threshold was found for outflowing clumps (\S~\ref{sec:non-static-case}). We also showed that for outflowing clumps, increasing $\fc$ naturally leads to more asymmetric line profiles, in much better agreement with what has been observed in observations of galaxies.
\end{itemize}
These results suggest important implications for the interpretation of observed \Lya spectra, as a multiphase medium is physically more motivated than simplified homogeneous geometries such as the `shell-model'. Nevertheless since shell models successfully reproduce observed spectra, they are frequently used to model observations. Because a medium with sufficiently large covering factor behaves as a homogeneous medium, the success of shell models may indicate 
large covering factors are typical in galaxies as predicted by \citet{McCourt2016}.
Specifically, we found typical values of $\fccrit \sim 10-50$, much smaller than $\fc \gtrsim 1000$ predicted in their work. In this picture, it is easy to understand the convergence to the shell model.

Motivated by these results, we fitted shell models to spectra emerging from extremely clumpy outflows undergoing Hubble flow. We found that the column density from the shell closely matches that of the collection of clumps as a whole, and the shell expansion velocity appears to be the mass weighted average velocity. This result is very promising as it suggests that the shell model provides us with a fast method of extracting some physical properties of the interstellar and circumgalactic medium from the \Lya spectral line shape.
In addition, the value of other shell parameters (e.g. intrinsic \Lya line width prior to scattering) should not necessarily be interpreted literally as physical.
We will explore this systematically in future work.

Another implication concerns the mismatch between observed \Lya spectra and the ones predicted by theoretical studies of \Lya radiative transfer utilizing hydrodynamical simulations for their input geometry. Our work suggests that this mismatch can be due to the existence of tiny clumps in the observed systems which cannot form even in the most modern hydrodynamical simulations of galaxies due to their limited resolution. Thus, setups of these simulations might yield effective covering factors which are too low causing the spectra to possess, e.g., a too large flux at line center.
We will use our results for radiative transfer on small scales to develop an effective theory which can be implemented as a sub-grid model in global simulations of galaxies.



\begin{acknowledgements}
This research made use of NASA's Astrophysics Data System, and a number of open source software such as the \texttt{IPython} package \citep{PER-GRA:2007}; \texttt{matplotlib} \citep{Hunter:2007}; and \texttt{SciPy} \citep{jones_scipy_2001}. MD and MG thank the astronomy group at UCSB for their hospitality, and the organizers and participants of `SnowCLAW 2017' for an inspiring conference.
MM was supported by NASA grant HST-HF2-51376.001-A, under NASA
contract NAS5-26555. SPO acknowledges NASA grant NNX15AK81G. 
\end{acknowledgements}

\appendix
\section{Regimes of a medium with uncorrelated clump motion}
\label{sec:regimes_moving}

\begin{figure}
  \centering
  \includegraphics[width=.95\linewidth]{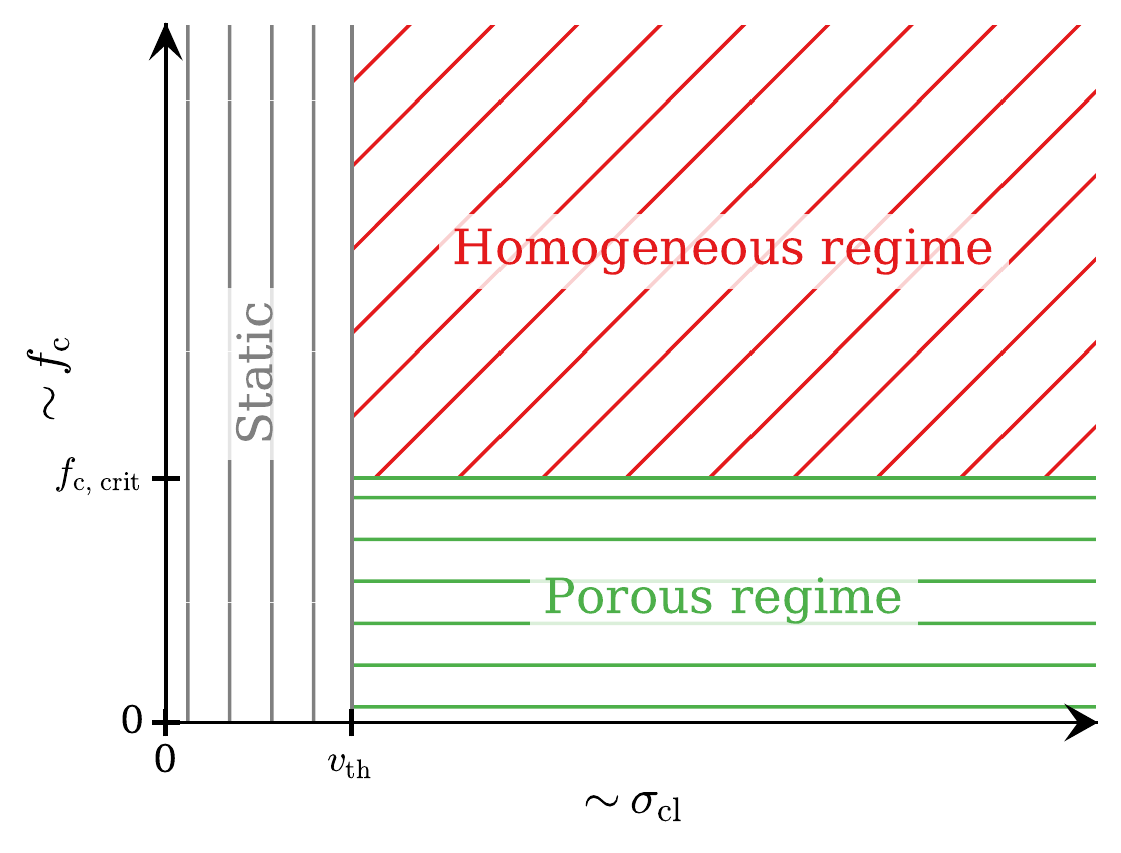}
  \caption{Escape regimes of a medium with (uncorrelated) randomly moving clumps as discussed in Appendix~\ref{sec:regimes_moving}.}
  \label{fig:regimes_sketch_moving}
\end{figure}

For randomly moving, optically thick clumps the photons either escape through holes in velocity space (if $\fc \lesssim \fccrit$, Eq.~\eqref{eq:fccrit_moving}) or escape in single-flight (for $\fc\gtrsim \fccrit$) -- as described in \S~\ref{sec:non-static-case}. In the former case, the emergent spectrum will be similar to the intrinsic one, that is, narrow, and singly peaked. If, however, $\fc > \fccrit$ the radiative transfer process will be similar to a slab with temperature $T_{\mathrm{eff}}$ (Eq.~\eqref{eq:Teff}) which means the photons will escape in a single flight after interaction with one (fast-moving) clump, and so will the emergent double peaked spectrum, i.e., a peak position of $x_{\rm p}\sim x_*$ or in (observed) velocity units
\begin{equation}
  \label{eq:v_peak_moving}
  v_{\mathrm{p}}\approx x_* v_{\mathrm{th}}(T_{\mathrm{eff}}) \approx 3.8\sigma_{\cl}\;.
\end{equation}
Fig.~\ref{fig:regimes_sketch_moving} shows a visual overview of these regimes. In this figure, we also marked that a smaller clump motion than the internal thermal motion of the atoms (for the parameters used in this work of $\sigma_\cl \lesssim 13\kms$) leads to a convergence back to the static case.

Another interesting part of the parameter space is between the two regimes, for $\fc \sim \fccrit$. Here, the velocity space is sufficiently sampled so that hardly any photons can escape without clump interaction. However, after an interaction with a (slowly moving) clump the probability of interacting with another clump is small -- even if the photon is still in the core of the line. This is because the velocity distribution is not \textit{that} well sampled to provide $\tau_0 \gg 1$.
As a result, the emergent spectrum will directly represent the clumps' velocity dispersion -- which means a single peaked spectrum a line center of width $\sim \sigma_\cl$.

To summarize: with increasing covering factor, a medium with uncorrelated clump motion can lead to a narrow or wide single peaked spectrum (of widths of the intrinsic spectrum or the clump velocity dispersion, respectively) at line center, or a wide double-peaked spectrum (if $\fc\gtrsim \fccrit$).

\section{Additional numerical results}
\label{sec:additional-results}

\begin{figure}
  \centering
  \includegraphics[width=.95\linewidth]{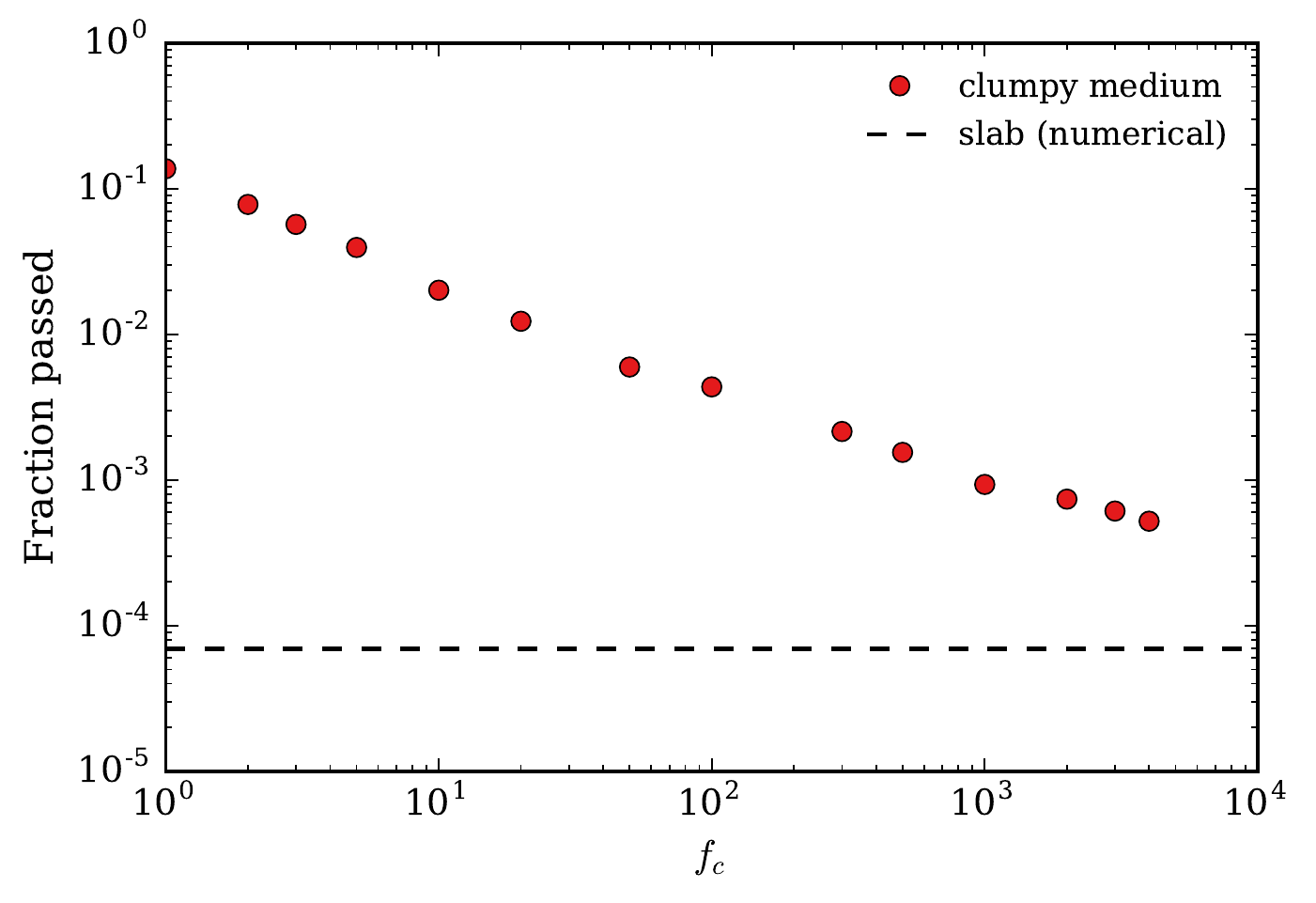}
  \caption{Transmission through a clumpy slab with column density $N_{\mathrm{HI,\,total}}=\frac{4}{3}\times 10^{19}\cm^{-2}$ (as before, measured per half-height) versus covering factor $\fc$. As comparison we show the transmission through a (solid) slab with the same column density.}
  \label{fig:transmission}
\end{figure}

\begin{figure*}
  \centering
  \includegraphics[width=.95\linewidth]{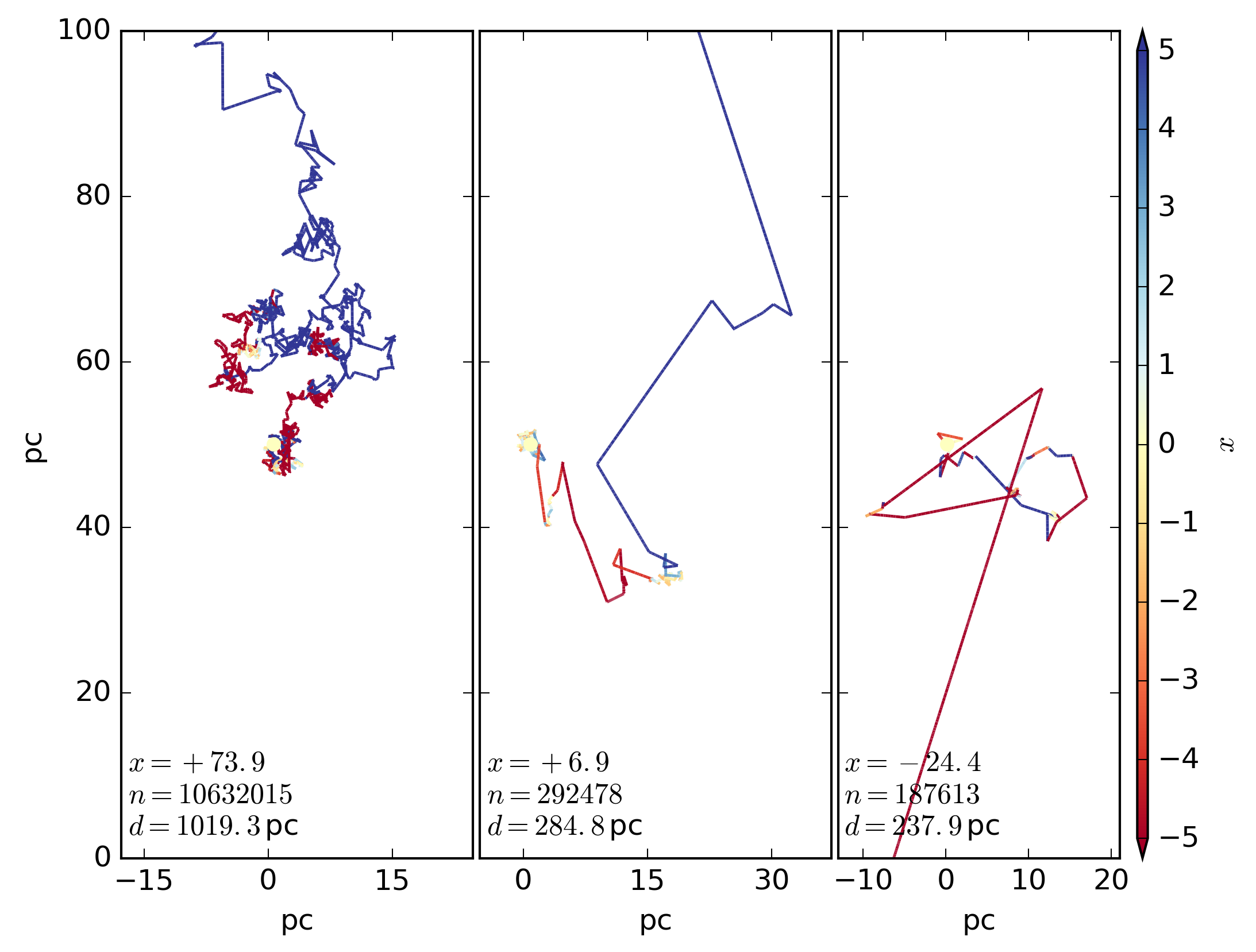}
  \caption{Examples of photon trajectories. The \textit{left panel} shows a photon escaping through random-walk from a static medium with $(N_{\HI,\cl},\,\fc) = (10^{20}\cm^{-2},\,100)$. In the \textit{central panel} the photon escapes in an excursion [$(N_{\HI,\cl},\,\fc) = (10^{17}\cm^{-2},\,100)$] after some random walk, and in the \textit{right panel} nearly directly through excursion / single flight due to movement of the clumps [$(\,N_{\HI,\cl},\,\fc,\sigma_\cl) = (10^{17}\cm^{-2},\,100,\,100\kms)$]. In each panel, the escape frequency $x$ is displayed as well as the total number of scatterings $n$, and the distance travelled $d$. In addition, the color coding represents the frequency of the photon (truncated at $\pm 5$). An animated version of this figure is available at \url{http://bit.ly/a-in-a-fog}. 
}
  \label{fig:example_traj}
\end{figure*}

\subsection{Transmission through a clumpy slab}
\label{sec:transm-thro-clumpy}
Fig.~\ref{fig:transmission} shows the fraction of photons that passed through a clumpy medium when emitted at the boundary of the box for a fixed total column density but various number of clumps per sightline. The transmitted fraction of photons decreases with increasing covering factor and approaches the limit of a homogeneous slab. We attribute this dependence on \fc as well as the differences to `surface effects', i.e., due to the roughness of the boundary it is easier for photons to get ``trapped'' in the slab.

\subsection{Examples of photon trajectories}
\label{sec:exampl-phot-traj}

Fig.~\ref{fig:example_traj} shows examples for the three different escape mechanism discussed in this work. The left panel shows a random walk in a static medium, the central panel escape through excursion, and the right panel the escape through single flight. An animated version of Fig.~\ref{fig:example_traj} is available online\footnote{\url{http://bit.ly/a-in-a-fog}}.


\bibliography{references}

\begin{thebibliography}{69}
\expandafter\ifx\csname natexlab\endcsname\relax\def\natexlab#1{#1}\fi

\bibitem[{Adams(1972)}]{Adams1972}
Adams, T.~F. 1972, \apj, 174, 439

\bibitem[{Adams(1975)}]{Adams1975}
Adams, T.~F. 1975, \apj, 201, 350

\bibitem[{Ahn {et~al.}(2002)Ahn, Lee, \& Lee}]{Ahn2001a}
Ahn, S.-H., Lee, H., \& Lee, H.~M. 2002, \apj, 567, 922

\bibitem[{{Ahn} {et~al.}(2003){Ahn}, {Lee}, \& {Lee}}]{Ahn2003MNRAS.340..863A}
{Ahn}, S.-H., {Lee}, H.-W., \& {Lee}, H.~M. 2003, \mnras, 340, 863

\bibitem[{{Arav} {et~al.}(1997){Arav}, {Barlow}, {Laor}, \&
  {Blandford}}]{1997MNRAS.288.1015A}
{Arav}, N., {Barlow}, T.~A., {Laor}, A., \& {Blandford}, R.~D. 1997, \mnras,
  288, 1015

\bibitem[{Auer(1968)}]{Auer1968}
Auer, L. H. P. U.~O. 1968, \apj, 153, pp.783

\bibitem[{{Bacon} {et~al.}(2010){Bacon}, {Accardo}, {Adjali}, {Anwand},
  {Bauer}, {Biswas}, {Blaizot}, {Boudon}, {Brau-Nogue}, {Brinchmann},
  {Caillier}, {Capoani}, {Carollo}, {Contini}, {Couderc}, {Daguis{\'e}},
  {Deiries}, {Delabre}, {Dreizler}, {Dubois}, {Dupieux}, {Dupuy}, {Emsellem},
  {Fechner}, {Fleischmann}, {Fran{\c c}ois}, {Gallou}, {Gharsa}, {Glindemann},
  {Gojak}, {Guiderdoni}, {Hansali}, {Hahn}, {Jarno}, {Kelz}, {Koehler},
  {Kosmalski}, {Laurent}, {Le Floch}, {Lilly}, {Lizon}, {Loupias}, {Manescau},
  {Monstein}, {Nicklas}, {Olaya}, {Pares}, {Pasquini}, {P{\'e}contal-Rousset},
  {Pell{\'o}}, {Petit}, {Popow}, {Reiss}, {Remillieux}, {Renault}, {Roth},
  {Rupprecht}, {Serre}, {Schaye}, {Soucail}, {Steinmetz}, {Streicher}, {Stuik},
  {Valentin}, {Vernet}, {Weilbacher}, {Wisotzki}, \&
  {Yerle}}]{2010SPIE.7735E..08B}
{Bacon}, R., {Accardo}, M., {Adjali}, L., {et~al.} 2010, in \procspie, Vol.
  7735, Ground-based and Airborne Instrumentation for Astronomy III, 773508

\bibitem[{Barnes {et~al.}(2014)Barnes, Garel, \& Kacprzak}]{Barnes2014}
Barnes, L.~A., Garel, T., \& Kacprzak, G.~G. 2014, Publications of the
  Astronomical Society of the Pacific, 126, 969

\bibitem[{{Barnes} {et~al.}(2011){Barnes}, {Haehnelt}, {Tescari}, \&
  {Viel}}]{2011MNRAS.416.1723B}
{Barnes}, L.~A., {Haehnelt}, M.~G., {Tescari}, E., \& {Viel}, M. 2011, \mnras,
  416, 1723

\bibitem[{Behrens \& Braun(2014)}]{Behrens2014a}
Behrens, C. \& Braun, H. 2014, \aap, 572, A74

\bibitem[{{Blanc} {et~al.}(2011){Blanc}, {Adams}, {Gebhardt}, {Hill}, {Drory},
  {Hao}, {Bender}, {Ciardullo}, {Finkelstein}, {Fry}, {Gawiser}, {Gronwall},
  {Hopp}, {Jeong}, {Kelzenberg}, {Komatsu}, {MacQueen}, {Murphy}, {Roth},
  {Schneider}, \& {Tufts}}]{2011ApJ...736...31B}
{Blanc}, G.~A., {Adams}, J.~J., {Gebhardt}, K., {et~al.} 2011, \apj, 736, 31

\bibitem[{{Bonilha} {et~al.}(1979){Bonilha}, {Ferch}, {Salpeter}, {Slater}, \&
  {Noerdlinger}}]{1979ApJ...233..649B}
{Bonilha}, J.~R.~M., {Ferch}, R., {Salpeter}, E.~E., {Slater}, G., \&
  {Noerdlinger}, P.~D. 1979, \apj, 233, 649

\bibitem[{{Cantalupo} {et~al.}(2014){Cantalupo}, {Arrigoni-Battaia},
  {Prochaska}, {Hennawi}, \& {Madau}}]{Cantalupo2014}
{Cantalupo}, S., {Arrigoni-Battaia}, F., {Prochaska}, J.~X., {Hennawi}, J.~F.,
  \& {Madau}, P. 2014, \nat, 506, 63

\bibitem[{Dijkstra(2014)}]{Dijkstra2014_review}
Dijkstra, M. 2014, Publications of the Astronomical Society of Australia, 31,
  26

\bibitem[{Dijkstra {et~al.}(2016)Dijkstra, Gronke, \&
  Venkatesan}]{DijkstraLyaLyC2016}
Dijkstra, M., Gronke, M., \& Venkatesan, A. 2016, \apj, 828, 71

\bibitem[{{Dijkstra} {et~al.}(2006){Dijkstra}, {Haiman}, \&
  {Spaans}}]{2006ApJ...649...14D}
{Dijkstra}, M., {Haiman}, Z., \& {Spaans}, M. 2006, \apj, 649, 14

\bibitem[{{Dijkstra} \& {Jeeson-Daniel}(2013)}]{2013MNRAS.435.3333D}
{Dijkstra}, M. \& {Jeeson-Daniel}, A. 2013, \mnras, 435, 3333

\bibitem[{{Dijkstra} \& {Kramer}(2012)}]{Dijksta2012MNRAS.424.1672D}
{Dijkstra}, M. \& {Kramer}, R. 2012, \mnras, 424, 1672

\bibitem[{Dijkstra {et~al.}(2007)Dijkstra, Wyithe, \& Haiman}]{Dijkstra2007}
Dijkstra, M., Wyithe, J. S.~B., \& Haiman, Z. 2007, \mnras, 379, 253

\bibitem[{Duval {et~al.}(2014)Duval, Schaerer, {\"{O}}stlin, \&
  Laursen}]{Duval2013}
Duval, F., Schaerer, D., {\"{O}}stlin, G., \& Laursen, P. 2014, \aap, 562, A52

\bibitem[{Erb {et~al.}(2014)Erb, Steidel, Trainor, Bogosavljevi{\'{c}},
  Shapley, Nestor, Kulas, Law, Strom, Rudie, Reddy, Pettini, Konidaris, Mace,
  Matthews, \& McLean}]{Erb2014}
Erb, D.~K., Steidel, C.~C., Trainor, R.~F., {et~al.} 2014, \apj, 795, 33

\bibitem[{{Faucher-Gigu{\`e}re} {et~al.}(2015){Faucher-Gigu{\`e}re}, {Hopkins},
  {Kere{\v s}}, {Muratov}, {Quataert}, \& {Murray}}]{2015MNRAS.449..987F}
{Faucher-Gigu{\`e}re}, C.-A., {Hopkins}, P.~F., {Kere{\v s}}, D., {et~al.}
  2015, \mnras, 449, 987

\bibitem[{Gronke {et~al.}(2015)Gronke, Bull, \& Dijkstra}]{Gronke2015}
Gronke, M., Bull, P., \& Dijkstra, M. 2015, \apj, 812, 123

\bibitem[{Gronke \& Dijkstra(2014)}]{Gronke2014a}
Gronke, M. \& Dijkstra, M. 2014, \mnras, 1103, 10

\bibitem[{Gronke \& Dijkstra(2016)}]{Gronke2016a}
Gronke, M. \& Dijkstra, M. 2016, \apj, 826, 14

\bibitem[{{Gronke} {et~al.}(2016){Gronke}, {Dijkstra}, {McCourt}, \&
  {Oh}}]{clumps1}
{Gronke}, M., {Dijkstra}, M., {McCourt}, M., \& {Oh}, S.~P. 2016, \apjl, 833,
  L26

\bibitem[{Hansen \& Oh(2006)}]{Hansen2005}
Hansen, M. \& Oh, S.~P. 2006, \mnras, 367, 979

\bibitem[{{Harrington}(1973)}]{1973MNRAS.162...43H}
{Harrington}, J.~P. 1973, \mnras, 162, 43

\bibitem[{Hashimoto {et~al.}(2015)Hashimoto, Verhamme, Ouchi, Shimasaku,
  Schaerer, Nakajima, Shibuya, Rauch, Ono, \& Goto}]{Hashimoto2015}
Hashimoto, T., Verhamme, A., Ouchi, M., {et~al.} 2015, \apj, 812, 157

\bibitem[{Hayes(2015)}]{Hayes2015}
Hayes, M. 2015, Publications of the Astronomical Society of Australia, 32, e027

\bibitem[{Hayes {et~al.}(2011)Hayes, Schaerer, {\"{O}}stlin, Mas-Hesse, Atek,
  \& Kunth}]{Hayes2011}
Hayes, M., Schaerer, D., {\"{O}}stlin, G., {et~al.} 2011, \apj, 730, 8

\bibitem[{{Hennawi} {et~al.}(2015){Hennawi}, {Prochaska}, {Cantalupo}, \&
  {Arrigoni-Battaia}}]{Hennawi2015}
{Hennawi}, J.~F., {Prochaska}, J.~X., {Cantalupo}, S., \& {Arrigoni-Battaia},
  F. 2015, Science, 348, 779

\bibitem[{Henry {et~al.}(2015)Henry, Scarlata, Martin, \& Erb}]{Henry2015}
Henry, A., Scarlata, C., Martin, C.~L., \& Erb, D. 2015, \apj, 809, 19

\bibitem[{Hunter(2007)}]{Hunter:2007}
Hunter, J.~D. 2007, Computing In Science \& Engineering, 9, 90

\bibitem[{Jones {et~al.}(2001)Jones, Oliphant, Peterson, \&
  Others}]{jones_scipy_2001}
Jones, E., Oliphant, T., Peterson, P., \& Others. 2001, {SciPy}: Open source
  scientific tools for Python

\bibitem[{{Karman} {et~al.}(2016){Karman}, {Caputi}, {Caminha}, {Gronke},
  {Grillo}, {Balestra}, {Rosati}, {Vanzella}, {Coe}, {Dijkstra}, {Koekemoer},
  {McLeod}, {Mercurio}, \& {Nonino}}]{Karman2016_dl}
{Karman}, W., {Caputi}, K.~I., {Caminha}, G.~B., {et~al.} 2016, preprint

\bibitem[{Kulas {et~al.}(2012)Kulas, Shapley, Kollmeier, Zheng, Steidel, \&
  Hainline}]{Kulas2011}
Kulas, K.~R., Shapley, A.~E., Kollmeier, J.~A., {et~al.} 2012, \apj, 745, 33

\bibitem[{Laursen {et~al.}(2013)Laursen, Duval, \& {\"{O}}stlin}]{Laursen2012}
Laursen, P., Duval, F., \& {\"{O}}stlin, G. 2013, \apj, 766, 124

\bibitem[{{Laursen} \& {Sommer-Larsen}(2007)}]{Laursen2007ApJ...657L..69L}
{Laursen}, P. \& {Sommer-Larsen}, J. 2007, \apjl, 657, L69

\bibitem[{Laursen {et~al.}(2009)Laursen, Sommer-Larsen, \&
  Andersen}]{Laursen2009}
Laursen, P., Sommer-Larsen, J., \& Andersen, A.~C. 2009, \apj, 704, 1640

\bibitem[{Laursen {et~al.}(2011)Laursen, Sommer-Larsen, \&
  Razoumov}]{Laursen2011}
Laursen, P., Sommer-Larsen, J., \& Razoumov, A.~O. 2011, \apj, 728, 52

\bibitem[{{Liang} {et~al.}(2016){Liang}, {Kravtsov}, \&
  {Agertz}}]{2016MNRAS.458.1164L}
{Liang}, C.~J., {Kravtsov}, A.~V., \& {Agertz}, O. 2016, \mnras, 458, 1164

\bibitem[{{Mas-Ribas} \& {Dijkstra}(2016)}]{Lluis2016ApJ...822...84M}
{Mas-Ribas}, L. \& {Dijkstra}, M. 2016, \apj, 822, 84

\bibitem[{McCourt {et~al.}(2016)McCourt, Oh, O'Leary, \& Madigan}]{McCourt2016}
McCourt, M., Oh, S.~P., O'Leary, R.~M., \& Madigan, A.-M. 2016, preprint
  [\eprint{arXiv:1610.01164}]

\bibitem[{McKee \& Ostriker(1977)}]{McKee1977}
McKee, C.~F. \& Ostriker, J.~P. 1977, \apj, 218, 148

\bibitem[{Neufeld(1990)}]{Neufeld1990}
Neufeld, D.~A. 1990, \apj, 350, 216

\bibitem[{Neufeld(1991)}]{Neufeld1991}
Neufeld, D.~A. 1991, \apj, 370, L85

\bibitem[{Osterbrock(1962)}]{Osterbrock1962}
Osterbrock, D.~E. 1962, \apj, 135, 195

\bibitem[{{\"{O}}stlin {et~al.}(2014){\"{O}}stlin, Hayes, Duval, Sandberg,
  Rivera-Thorsen, Marquart, Orlitov{\'{a}}, Adamo, Melinder, Guaita, Atek,
  Cannon, Gruyters, Herenz, Kunth, Laursen, Mas-Hesse, Micheva,
  Ot{\'{i}}-Floranes, Pardy, Roth, Schaerer, \& Verhamme}]{Ostlin2014}
{\"{O}}stlin, G., Hayes, M., Duval, F., {et~al.} 2014, \apj, 797, 11

\bibitem[{Pei(1992)}]{Pei1992}
Pei, Y.~C. 1992, \apj, 395, 130

\bibitem[{P\'erez \& Granger(2007)}]{PER-GRA:2007}
P\'erez, F. \& Granger, B.~E. 2007, Computing in Science and Engineering, 9, 21

\bibitem[{{Rauch} {et~al.}(1999){Rauch}, {Sargent}, \& {Barlow}}]{Rauch1999}
{Rauch}, M., {Sargent}, W.~L.~W., \& {Barlow}, T.~A. 1999, \apj, 515, 500

\bibitem[{Rees(1987)}]{Rees1987}
Rees, M.~J. 1987, \mnras, 228, 47P

\bibitem[{{Rivera-Thorsen} {et~al.}(2015){Rivera-Thorsen}, {Hayes},
  {{\"O}stlin}, {Duval}, {Orlitov{\'a}}, {Verhamme}, {Mas-Hesse}, {Schaerer},
  {Cannon}, {Ot{\'{\i}}-Floranes}, {Sandberg}, {Guaita}, {Adamo}, {Atek},
  {Herenz}, {Kunth}, {Laursen}, \& {Melinder}}]{Rivera-thorsen2014}
{Rivera-Thorsen}, T.~E., {Hayes}, M., {{\"O}stlin}, G., {et~al.} 2015, \apj,
  805, 14

\bibitem[{Smith {et~al.}(2015)Smith, Safranek-Shrader, Bromm, \&
  Milosavljevi}]{Smith2014}
Smith, A., Safranek-Shrader, C., Bromm, V., \& Milosavljevi, M. 2015, \mnras,
  449, 4336

\bibitem[{{Song} {et~al.}(2014){Song}, {Finkelstein}, {Gebhardt}, {Hill},
  {Drory}, {Ashby}, {Blanc}, {Bridge}, {Chonis}, {Ciardullo}, {Fabricius},
  {Fazio}, {Gawiser}, {Gronwall}, {Hagen}, {Huang}, {Jogee}, {Livermore},
  {Salmon}, {Schneider}, {Willner}, \& {Zeimann}}]{2014ApJ...791....3S}
{Song}, M., {Finkelstein}, S.~L., {Gebhardt}, K., {et~al.} 2014, \apj, 791, 3

\bibitem[{Steidel {et~al.}(2010)Steidel, Erb, Shapley, Pettini, Reddy,
  Bogosavljevi{\'{c}}, Rudie, \& Rakic}]{Steidel2010a}
Steidel, C.~C., Erb, D.~K., Shapley, A.~E., {et~al.} 2010, \apj, 717, 289

\bibitem[{{Steidel} {et~al.}(2014){Steidel}, {Rudie}, {Strom}, {Pettini},
  {Reddy}, {Shapley}, {Trainor}, {Erb}, {Turner}, {Konidaris}, {Kulas}, {Mace},
  {Matthews}, \& {McLean}}]{2014ApJ...795..165S}
{Steidel}, C.~C., {Rudie}, G.~C., {Strom}, A.~L., {et~al.} 2014, \apj, 795, 165

\bibitem[{{Tasitsiomi}(2006)}]{2006ApJ...645..792T}
{Tasitsiomi}, A. 2006, \apj, 645, 792

\bibitem[{Trainor {et~al.}(2015)Trainor, Steidel, Strom, \&
  Rudie}]{Trainor2015}
Trainor, R.~F., Steidel, C.~C., Strom, A.~L., \& Rudie, G.~C. 2015, \apj, 809,
  89

\bibitem[{Trebitsch {et~al.}(2016)Trebitsch, Verhamme, Blaizot, \&
  Rosdahl}]{Trebitsch2016}
Trebitsch, M., Verhamme, A., Blaizot, J., \& Rosdahl, J. 2016, \aap, 593, A122

\bibitem[{Verhamme {et~al.}(2012)Verhamme, Dubois, Blaizot, Garel, Bacon,
  Devriendt, Guiderdoni, \& Slyz}]{Verhamme2012}
Verhamme, A., Dubois, Y., Blaizot, J., {et~al.} 2012, \aap, 546, A111

\bibitem[{{Verhamme} {et~al.}(2015){Verhamme}, {Orlitov{\'a}}, {Schaerer}, \&
  {Hayes}}]{2015A&A...578A...7V}
{Verhamme}, A., {Orlitov{\'a}}, I., {Schaerer}, D., \& {Hayes}, M. 2015, \aap,
  578, A7

\bibitem[{{Verhamme} {et~al.}(2006){Verhamme}, {Schaerer}, \&
  {Maselli}}]{Verhamme2006A&A...460..397V}
{Verhamme}, A., {Schaerer}, D., \& {Maselli}, A. 2006, \aap, 460, 397

\bibitem[{{Yamada} {et~al.}(2012){Yamada}, {Matsuda}, {Kousai}, {Hayashino},
  {Morimoto}, \& {Umemura}}]{2012ApJ...751...29Y}
{Yamada}, T., {Matsuda}, Y., {Kousai}, K., {et~al.} 2012, \apj, 751, 29

\bibitem[{Yang {et~al.}(2016)Yang, Malhotra, Gronke, Rhoads, Dijkstra, Jaskot,
  Zheng, \& Wang}]{Yang2015}
Yang, H., Malhotra, S., Gronke, M., {et~al.} 2016, \apj, 820, 130

\bibitem[{Yang {et~al.}(2017)Yang, Malhotra, Gronke, Rhoads, Leitherer,
  Wofford, Jiang, Dijkstra, Tilvi, \& Wang}]{Yang2017}
Yang, H., Malhotra, S., Gronke, M., {et~al.} 2017, preprint
  [\eprint{arXiv:1701.01857}]

\bibitem[{Zheng {et~al.}(2010)Zheng, Cen, Trac, \&
  Miralda-Escud{\'{e}}}]{Zheng2009}
Zheng, Z., Cen, R., Trac, H., \& Miralda-Escud{\'{e}}, J. 2010, \apj, 716, 574

\bibitem[{Zheng \& Miralda-Escud{\'{e}}(2002)}]{Zheng2002}
Zheng, Z. \& Miralda-Escud{\'{e}}, J. 2002, \apj, 578, 33

\end{thebibliography}

\end{document}